\makeatletter
\declare@file@substitution{revtex4-1.cls}{revtex4-2.cls}
\makeatother
\documentclass[twocolumn]{aastex631}

\usepackage{array,multirow}

\shorttitle{Contribution of velocity dispersion in molecular clouds}
\shortauthors{Yuan et al.}
\graphicspath{{./}{figures/}}

\begin{document}

\title{Understanding the Kinetic Energy deposition within Molecular Clouds}

\correspondingauthor{Ji Yang}
\email{jiyang@pmo.ac.cn}

\author[0000-0003-0804-9055]{Lixia Yuan}
\affiliation{Purple Mountain Observatory and Key Laboratory of Radio Astronomy, Chinese Academy of Sciences, \\
10 Yuanhua Road, Qixia District, Nanjing 210033, PR China}
\email{lxyuan@pmo.ac.cn}

\author[0000-0001-7768-7320]{Ji Yang}
\affiliation{Purple Mountain Observatory and Key Laboratory of Radio Astronomy, Chinese Academy of Sciences, \\
10 Yuanhua Road, Qixia District, Nanjing 210033, PR China}

\author[0000-0002-7489-0179]{Fujun Du}
\affiliation{Purple Mountain Observatory and Key Laboratory of Radio Astronomy, Chinese Academy of Sciences, \\
10 Yuanhua Road, Qixia District, Nanjing 210033, PR China}

\author[0000-0002-0197-470X]{Yang Su}
\affiliation{Purple Mountain Observatory and Key Laboratory of Radio Astronomy, Chinese Academy of Sciences, \\
10 Yuanhua Road, Qixia District, Nanjing 210033, PR China}

\author[0000-0003-2549-7247]{Shaobo Zhang}
\affiliation{Purple Mountain Observatory and Key Laboratory of Radio Astronomy, Chinese Academy of Sciences, \\
10 Yuanhua Road, Qixia District, Nanjing 210033, PR China}

\author[0000-0003-4586-7751]{Qing-Zeng Yan}
\affiliation{Purple Mountain Observatory and Key Laboratory of Radio Astronomy, Chinese Academy of Sciences, \\
10 Yuanhua Road, Qixia District, Nanjing 210033, PR China}

\author[0000-0002-3904-1622]{Yan Sun}
\affiliation{Purple Mountain Observatory and Key Laboratory of Radio Astronomy, Chinese Academy of Sciences, \\
10 Yuanhua Road, Qixia District, Nanjing 210033, PR China}

\author[0000-0003-2418-3350]{Xin Zhou}
\affiliation{Purple Mountain Observatory and Key Laboratory of Radio Astronomy, Chinese Academy of Sciences, \\
10 Yuanhua Road, Qixia District, Nanjing 210033, PR China}

\author[0000-0003-3151-8964]{Xuepeng Chen}
\affiliation{Purple Mountain Observatory and Key Laboratory of Radio Astronomy, Chinese Academy of Sciences, \\
10 Yuanhua Road, Qixia District, Nanjing 210033, PR China}

\author[0000-0003-0746-7968]{Hongchi Wang}
\affiliation{Purple Mountain Observatory and Key Laboratory of Radio Astronomy, Chinese Academy of Sciences, \\
10 Yuanhua Road, Qixia District, Nanjing 210033, PR China}

\author[0000-0003-0849-0692]{Zhiwei Chen}
\affiliation{Purple Mountain Observatory and Key Laboratory of Radio Astronomy, Chinese Academy of Sciences, \\
10 Yuanhua Road, Qixia District, Nanjing 210033, PR China} 

\begin{abstract}
According to the structures traced by $^{13}$CO spectral lines within the $^{12}$CO molecular clouds (MCs), 
we investigate the contributions of their internal gas motions and relative motions 
to the total velocity dispersions of $^{12}$CO MCs. 
Our samples of 2851 $^{12}$CO MCs harbor a total of 9556 individual $^{13}$CO structures, 
among which 1848 MCs ($\sim$ 65$\%$) have one individual $^{13}$CO structure and the other 1003 MCs ($\sim$ 35$\%$) 
have multiple $^{13}$CO structures. 
We find that the contribution of the relative motion between $^{13}$CO structures ($\sigma_{\rm ^{13}CO, re}$) 
is larger than that from their internal gas motion ($\sigma_{\rm ^{13}CO, in}$) in $\sim$ 62$\%$ of 1003 MCs in the `multiple' regime.
In addition, we find the $\sigma_{\rm ^{13}CO, re}$ tends to increase with the total velocity dispersion($\sigma_{\rm ^{12}CO, tot}$) in our samples, 
especially for the MCs having multiple $^{13}$CO structures. 
This result provides a manifestation of the macro-turbulent within MCs, 
which gradually becomes the dominant way to store the kinetic energy along with the development of MC scales.

\end{abstract}


\section{Introduction}\label{sec:intro}

Molecular clouds (MCs) are the fundamental components within galaxies and also the sites of star formation.
Thus understanding the dynamic evolution of MCs is crucial for 
understanding how the large-scale gas gradually gathers into the stellar system. 
Several scenarios of the formation and evolution of MCs have been proposed, 
mainly including the top-down, bottom-up, and transient pictures. 
The top-down picture depicts that the large-scale clouds monotonously fragment into a hierarchy of successively small-scale clouds as the 
density rises and the Jeans mass decreases, due to the gravitational instabilities \citep{Oort1954, Lin1964, Goldreich1965}. 
The bottom-up models are the agglomeration or coagulation between smaller clouds to construct larger 
clouds monotonously \citep{Field1965, Kwan1983, Kwan1987, Tomisaka1984, Tomisaka1986}. 
In the transient picture, MCs formed in the diffuse ISM through local compressions induced by 
the different scales of converging flows, appear to evolve in either direction 
\citep{Vazquez1995, Passot1995, Ballesteros1999, Vazquez2006, Heitsch2006, Beuther2020}. 
Other interesting theories on the formation and evolution of MCs also have been discussed \citep[e.g.][]{Dobbs2014, Ballesteros2020, Chevance2023}.
Note that these mechanisms are not necessarily mutually exclusive and may 
have a role at various stages of the dynamical evolution of MCs \citep{Dobbs2008,Tasker2009, Dobbs2013, Dobbs2015, Ballesteros2020, Jeffreson2021}.  
However, the observational supports for these scenarios are still insufficient. 
Revealing the gas kinematics within MCs on scales ranging from MCs to their internal substructures 
are able to provide more observational constrains to the mechanisms acting on the MCs evolution.

One of the most fundamental and influential descriptions on the dynamical states of MCs
is parameterized by the Larson relations in \cite{Larson1981}, 
which is the power-law relationship between the global velocity dispersions ($\delta$V/km s$^{-1}$) and spatial sizes (L/pc) of MCs. 
The $\delta$V-L relation was investigated in different scales from the cloud-to-cloud \citep{Larson1981, Solomon1987, Heyer2009}, 
cloud-to-subregion \citep{Myers1983}, to the velocity structure functions in individual clouds \citep{Heyer2004, Heyer2006, Brunt2009}. 
However, the velocity dispersions of MCs in these relations are their total velocity dispersions, 
which are induced by the thermal and nonthermal gas motion. 
Taking the complex environment of the Galaxy into account, the molecular gas motion can be tied to 
the large-scale galactic-dynamical processes, e.g., the differential rotation and shear \citep{Bonnell2006}, 
and the stellar feedback, e.g., the expansion of HII regions or supernova explosions \citep{Skarbinski2023}.  
The nonthermal motions are composed of systematic and turbulent gas motion, 
and the turbulent motion contains both microturbulent and macroturbulent \citep{Zuckerman1974, Silk1985, Hacar2016}. 
We still lack information on the contributions of these different motions to the total velocity dispersion. 

Our previous work in \cite{Yuan2022} (Paper II) has identified the $^{13}$CO structures within 2851 $^{12}$CO MCs 
using the CO lines data from the Milky Way Imaging Scroll Painting (MWISP) survey \citep{Su2019}. 
We further investigated the spatial distribution of these $^{13}$CO structures, 
and found that there is a preferred spatial separation between $^{13}$CO structures in these $^{12}$CO clouds \citep{Yuan2023} (Paper III). 
In addition, a scaling relation between the $^{12}$CO cloud area and its harbored $^{13}$CO structure count 
also has been revealed. According to these observational results, we propose an alternative picture for 
the dynamical evolution of MCs: the regularly-spaced $^{13}$CO structures as the fundamental units build up the MCs, 
and the assembly and destruction of MCs proceeds under this fundamental unit. 
This picture implies the contribution of the relative motion between $^{13}$CO structures in the 
total velocity structures should not be ignorable. 
Meanwhile, the decomposition of the gas motion within MCs is essential to verify this picture. 

In this paper, we aim to decompose the total velocity dispersion into different components, 
which are mainly from the internal and relative movements of the $^{13}$CO structures within MCs. 
Section 2 mainly describes the $^{12}$CO and $^{13}$CO lines data from the MWISP survey, the 
extracted $^{12}$CO MC samples and their internal $^{13}$CO structures. 
In section 3, we present the results, including the systematic velocities 
derived by the $^{12}$CO and $^{13}$CO line emission for each MC, 
the distributions of the total velocity dispersion and its components for each MC, 
and the correlations between the total velocity dispersion and its components. 
In section 4, we discuss the effects of the derived systematic velocities 
and the optical depths of $^{12}$CO lines on our results. 
Meanwhile, we speculate about the micro-turbulent and macro-turbulent motions within MCs 
and also provide the implication on the dynamic evolution of MCs. 

\section{Data}
\subsection{$^{12}$CO(J=1-0) and $^{13}$CO(J=1-0) spectral lines data from the MWISP survey} 
The $^{12}$CO and $^{13}$CO at $J$=1-0 transition lines are from the Milky Way Imaging Scroll Painting (MWISP) survey, 
which is an ongoing northern Galactic plane CO survey. 
This CO survey is carried out on the 13.7m telescope at Delingha, China, and observes $^{12}$CO, 
$^{13}$CO, and C$^{18}$O($J$=1-0) lines simultaneously. 
A detailed description on the performance of the telescope and its multibeam receiver system is given 
in \cite{Su2019, Shan2012}.  
The observational strategy and the raw data reduction are also introduced in \cite{Su2019}.
The half-power beamwidth (HPBW) is $\sim$ 50$^{\prime \prime}$ at 115 GHz. 
The typical noise temperature is $\sim$ 250 K at $^{12}$CO lines and $\sim$ 140 K at $^{13}$CO and C$^{18}$O lines. 
A velocity resolution is about 0.16 km s$^{-1}$ for $^{12}$CO line and 0.17 km s$^{-1}$ for $^{13}$CO and C$^{18}$O lines, 
with a typical rms level of $\sim$ 0.5 K for $^{12}$CO lines and $\sim$ 0.3 K for the $^{13}$CO lines, respectively. 

In this work, we utilize the $^{12}$CO and $^{13}$CO line emission in the second Galactic quadrant 
with the Galactical longitude from  104$^{\circ}$.75 to 150$^{\circ}$.25, the Galactical latitude $|b| < 5^{\circ}.25$, 
and the velocity of $-$95 km s$^{-1}$  $<$ V$_{\rm LSR}$ $<$ 25 km s$^{-1}$.  
These $^{12}$CO and $^{13}$CO lines emission data also have been analyzed in \cite{Yuan2021, Yuan2022, Yuan2023}. 
 
\subsection{The $^{12}$CO molecular clouds and $^{13}$CO structures}
The $^{12}$CO molecular cloud in this work is defined as a set of contiguous voxels in the position-position-velocity (PPV) space 
with $^{12}$CO(1-0) line intensities above a certain threshold.  
A catalog of 18,190 $^{12}$CO molecular clouds has been identified from the $^{12}$CO line emission in the above region, 
using the Density-based Spatial Clustering of Applications with Noise (DBSCAN) algorithm \citep{ester1996, Yan2021}.  
The DBSCAN algorithm was designed to discover clusters in arbitrary shapes \citep{ester1996}, 
and further developed to identify the $^{12}$CO MCs by \cite{Yan2020}. 
This method combines both intensity levels and connectivity of signals to extract structures, 
which is fit for the extended and irregular shapes of MCs. 
These 18,190 MCs were visually inspected and also took a morphological classification, 
which was mainly classified into filaments and nonfilaments in \cite{Yuan2021} (Paper I). 

An individual $^{13}$CO structure in this work is characterized by a set of connected voxels in the PPV space having $^{13}$CO 
line intensities above a certain threshold, which is extracted using the DBSCAN algorithm within the boundaries of $^{12}$CO clouds. 
Among the total 18,190 $^{12}$CO clouds, 2851 $^{12}$CO clouds are identified to have $^{13}$CO structures \citep{Yuan2022}. 
The properties of these extracted $^{13}$CO structures have been systematically analyzed in Papers II and III.
The performance of the DBSCAN algorithm and the parameters used for the extraction of $^{12}$CO clouds 
and $^{13}$CO structures are described in detail in Appendix A. 
The extracted $^{12}$CO line data for 18,190 $^{12}$CO clouds and the extracted 
$^{13}$CO line data within the 2851 $^{12}$CO clouds are available at DOI:\href{https://doi.org/10.57760/sciencedb.j00001.00427}{10.57760/sciencedb.j00001.00427}. 
In this work, we focus on these 2851 $^{12}$CO molecular clouds and their internal $^{13}$CO structures.

\section{Results}
The goal of this work is to decompose the contributions of the internal and relative gas motion to 
the total velocity dispersion of a MC, according to the gas movements of its internal $^{13}$CO structures. 
Our samples of 2851 MCs harbor a total of 9566 individual $^{13}$CO structures, 
among which 1848 MCs (65$\%$) have one $^{13}$CO structure and the other 1003 MCs (35$\%$) have more than one $^{13}$CO structure. 
According to this, the whole clouds are separated into two regimes, i.e. single and multiple.
In a cloud, we define its internal region having both $^{12}$CO and $^{13}$CO emission as the $^{13}$CO-bright region, 
which contains the whole individual $^{13}$CO structures within a MC,
and the region with the $^{12}$CO line emission but not the $^{13}$CO emission as the $^{13}$CO-dark region.  

\subsection{Systematical velocities of $^{12}$CO molecular clouds}
To calculate the total velocity dispersion of a MC, its systematic velocity due to the differential rotation 
of the Galaxy needs to be determined first. 
According to the $^{12}$CO and $^{13}$CO line emission of MCs, we can calculate their centroid velocities to 
represent the systematic velocities of MCs. 
Using the $^{12}$CO and $^{13}$CO emission in a position-position-velocity (PPV) space for each cloud, 
we derive the centroid velocity for a MC as follows \citep{Rosolowsky2006}:
\begin{equation} \label{e:cen_12}
V_{\rm cen,^{12}CO}= \Sigma^{\rm cloud}_{i} T_{\rm ^{12}CO,i} V_{\rm ^{12}CO,i}/\Sigma^{\rm cloud}_{i} T_{\rm ^{12}CO,i},
\end{equation}

\begin{equation} \label{e:cen_13}
V_{\rm cen,^{13}CO}= \Sigma^{\rm ^{13}CO-bri}_{i} T_{\rm ^{13}CO,i} V_{\rm ^{13}CO,i}/\Sigma^{\rm ^{13}CO-bri}_{i} T_{\rm ^{13}CO,i},
\end{equation}
where the V$_{\rm ^{12}CO, i}$ and T$_{\rm ^{12}CO, i}$ are the line-of-sight velocity and brightness temperature 
of $^{12}$CO emission at the $i$th voxel in the PPV space of the whole $^{12}$CO cloud, 
the V$_{\rm ^{13}CO, i}$ and T$_{\rm ^{13}CO, i}$ are those values of the $^{13}$CO emission in the $^{13}$CO-bright region of a MC, 
the sum $\Sigma^{\rm cloud}_{i}$ runs over all voxels within the PPV space of a $^{12}$CO cloud, 
the sum $\Sigma^{\rm ^{13}CO-bri}_{i}$ runs over all voxels within the $^{13}$CO-bright region of a $^{12}$CO cloud.

\begin{figure*}[ht]
    \plotone{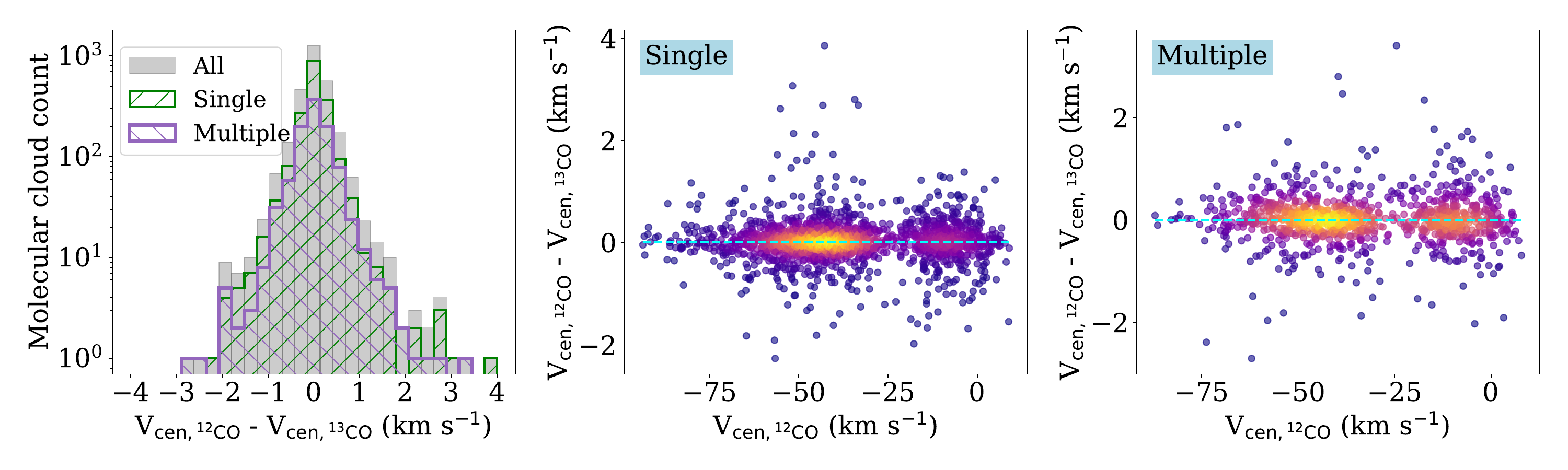} 
    \caption{The distributions of the differences between the centroid velocities from $^{12}$CO 
    emission ($V_{\rm cen,^{12}CO}$) and $^{13}$CO emission ($V_{\rm cen,^{13}CO}$) in MCs. 
    Among that, the `All' represents the whole 2851 MCs, and the `Single' and `Multiple' correspond to 
    the MCs having single and multiple (more than one) $^{13}$CO structures, respectively. 
    In the middle and right panels, each dot represents a $^{12}$CO MC. 
    The colors on these dots represent the distribution of the probability density function (2D-PDF) of 
    their $^{12}$CO MCs, which are calculated utilizing the kernel-density estimation through Gaussian kernels in the PYTHON package of 
    \href{https://docs.scipy.org/doc/scipy/reference/generated/scipy.stats.gaussian\_kde.html}{scipy.stats.gaussian\_kde}. 
    The cyan-dashed lines indicate the lines with $V_{\rm cen,^{12}CO}$ = $V_{\rm cen,^{13}CO}$. \label{fig:f_cenvel}}
\end{figure*}

Figure \ref{fig:f_cenvel} shows the differences between the centroid velocities calculated by 
the $^{12}$CO and $^{13}$CO emissions for each cloud. 
The quantiles of the differences between $V_{\rm cen,^{12}CO}$ and $V_{\rm cen,^{13}CO}$ for 
the MCs in the `all', `single', and `multiple' regimes are listed in Table \ref{tab:t_vcen}, respectively. 
We find that the differences between $V_{\rm cen,^{12}CO}$ and $V_{\rm cen,^{13}CO}$ are less than $\sim$ 0.65 km s$^{-1}$ 
for 90$\%$ of all the samples and less than $\sim$ 0.15 km s$^{-1}$ for 50$\%$ of the samples. 
The differences between $V_{\rm cen,^{12}CO}$ and $V_{\rm cen,^{13}CO}$ for the MCs in the `multiple' regime 
are a bit more scattered than those in the MCs in the `single' regime. 
The differences for 50$\%$ of MCs in the `multiple' regime are within $\sim$ 0.2 km s$^{-1}$, 
while this value is $\sim$ 0.15 km s$^{-1}$ for the MCs in the `single'. 
The median values are close to zero for the MCs with either single or multiple $^{13}$CO structures. 
Overall, the differences between $V_{\rm cen,^{12}CO}$ and $V_{\rm cen,^{13}CO}$ for most MCs concentrate 
in a narrow range, especially compared with the $^{12}$CO velocity span of MCs with a median value of 4.0 km s$^{-1}$ \citep{Yuan2022}.
The centroid velocity represents the systematical motion of the total gas in a single MC, 
although $^{12}$CO emission is easy to be optically thick, while it can trace most gas in a MC. 
Thus we first take the $V_{\rm cen,^{12}CO}$ as the systematic velocity ($V_{\rm sys}$) of MCs in the 
following calculations. The effects of the $V_{\rm cen,^{12}CO}$ or $V_{\rm cen,^{13}CO}$ as the systematic velocities ($V_{\rm sys}$) of 
MCs on the results also have been analyzed in the discussion. 

\begin{deluxetable*}{lccccc}
    \tablenum{1}
    \tablecaption{Differences between centroid velocities (km s$^{-1}$) from $^{12}$CO and $^{13}$CO line emission in MCs \label{tab:t_vcen}}
    \tablewidth{0pt}
    \tablehead{\colhead{Samples} & \colhead{0.05} & \colhead{0.25} &
    \colhead{0.5} & \colhead{0.75} & \colhead{0.95}\\
    }
    \startdata
    All & -0.64 & -0.14 & 0.01 & 0.18 & 0.66  \\
    Single & -0.58 & -0.12 & 0.02 & 0.17 & 0.60 \\
    Multiple & -0.69 & -0.20 & 0.0 & 0.2 & 0.70 \\
    \enddata
    \tablecomments{The quantiles at 0.05, 0.25, 0.5, 0.75, and 0.95 for the differences between $V_{\rm cen,^{12}CO}$ (km s$^{-1}$) and $V_{\rm cen,^{13}CO}$ (km s$^{-1}$) in their sequential data.
    The `All' represent the whole 2851 MCs, and the `Single' and `Multiple' correspond to the MCs having single and multiple $^{13}$CO structures, respectively.}
\end{deluxetable*}

\subsection{Decomposition of the total velocity dispersion in each MC}
Furthermore, we calculate the total velocity dispersion and its compositions from the gas motions in each MC.
The total velocity dispersion of a MC ($\sigma_{\rm ^{12}CO,tot}$) is calculated using its $^{12}$CO line emission as:
\begin{equation}
    \sigma_{\rm ^{12}CO,tot}^{2} = \Sigma^{\rm cloud}_{i} T_{\rm ^{12}CO,i} (V_{\rm ^{12}CO,i}-V_{\rm sys})^{2}/\Sigma^{\rm cloud}_{i} T_{\rm ^{12}CO,i},
\end{equation}
where the sum $\Sigma^{\rm cloud}_{i}$ runs over all voxels within the PPV space of a $^{12}$CO cloud.

The velocity dispersion for the gas in the $^{13}$CO-bright region ($\sigma_{\rm ^{13}CO,tot}$) is calculated as:
\begin{equation}
    \sigma_{\rm ^{13}CO,tot}^{2} = \Sigma^{\rm ^{13}CO-bri}_{i} T_{\rm ^{13}CO,i} (V_{\rm ^{13}CO,i}-V_{\rm sys})^{2}/\Sigma^{\rm ^{13}CO-bri}_{i} T_{\rm ^{13}CO,i},
\end{equation}
where the sum $\Sigma^{\rm ^{13}CO-bri}_{i}$ runs over all voxels within the $^{13}$CO-bright region of a cloud.

For a cloud, it is made up of the $^{13}$CO-bright and the $^{13}$CO-dark regions. 
The $^{13}$CO-bright region includes a single or multiple (more than one) individual $^{13}$CO structures, 
thus $\sigma_{\rm ^{13}CO,tot}$ can be decomposed into the internal gas motion within $^{13}$CO structures 
and the relative motion between $^{13}$CO structures.
Thus, for a cloud having $^{13}$CO structures with the number of $j$, the $\sigma_{\rm ^{13}CO,tot}$ can be further decomposed as:

\begin{eqnarray} \label{e:decom}
    \sigma_{\rm ^{13}CO,tot}^{2} & = & \sigma_{\rm ^{13}CO,in}^{2} + \sigma_{\rm ^{13}CO,re}^{2}, \nonumber\\
    \sigma_{\rm ^{13}CO,in}^{2} & = & \Sigma^{\rm cloud}_{j} F_{\rm ^{13}CO,j} \sigma_{\rm ^{13}CO,j}^{2} /\Sigma^{\rm cloud}_{j} F_{\rm ^{13}CO,j}, \nonumber \\
    \sigma_{\rm ^{13}CO,j}^{2} & = & \Sigma^{j\rm th}_{i} T_{\rm ^{13}CO,ji}(V_{\rm ^{13}CO,ji}-V_{\rm cen,^{13}CO, j})^{2}/\Sigma^{j\rm th}_{i} T_{\rm ^{13}CO,ji}, \nonumber \\
    V_{\rm cen,^{13}CO,j} & = & \Sigma^{j\rm th}_{i} T_{\rm ^{13}CO,ji} V_{\rm ^{13}CO,ji}/\Sigma^{j\rm th}_{i} T_{\rm ^{13}CO,ji}, \nonumber \\
    \sigma_{\rm ^{13}CO,re}^{2} & = & \Sigma^{\rm cloud}_{j} F_{\rm ^{13}CO,j} (V_{\rm cen,^{13}CO,j} - V_{\rm sys})^{2}/\Sigma^{\rm cloud}_{j} F_{\rm ^{13}CO,j}, \nonumber \\
\end{eqnarray}

where the $\sigma_{\rm ^{13}CO,in}$ represents the internal gas motion within the $^{13}$CO structures in a $^{12}$CO cloud, 
the $\sigma_{\rm ^{13}CO,re}$ represents the relative motion between the $^{13}$CO structures in a $^{12}$CO cloud.  
The $\sigma_{\rm ^{13}CO,j}$ is the velocity dispersion within the $j$th $^{13}$CO structure,
the $V_{\rm cen,^{13}CO,j}$ is the centroid velocity of $^{13}$CO emission in the $j$th $^{13}$CO structure,
the $T_{\rm ^{13}CO,ji}$ and $V_{\rm ^{13}CO,ji}$ are the brightness temperature and line-of-sight velocity of 
$^{13}$CO emission at the $i$th voxel in the $j$th $^{13}$CO structure. 
Among that, $F_{\rm ^{13}CO, j}=\int T_{mb}(l,b,v)dldbdv=0.167 \times 0.25 \Sigma_{i}^{jth} T_{\rm ^{13}CO,ji}$ (K km s$^{-1}$ arcmin$^{2}$) 
is the integrated flux of $^{13}$CO line emission for the $j$th $^{13}$CO structure.

\begin{figure*}[ht]
\plotone{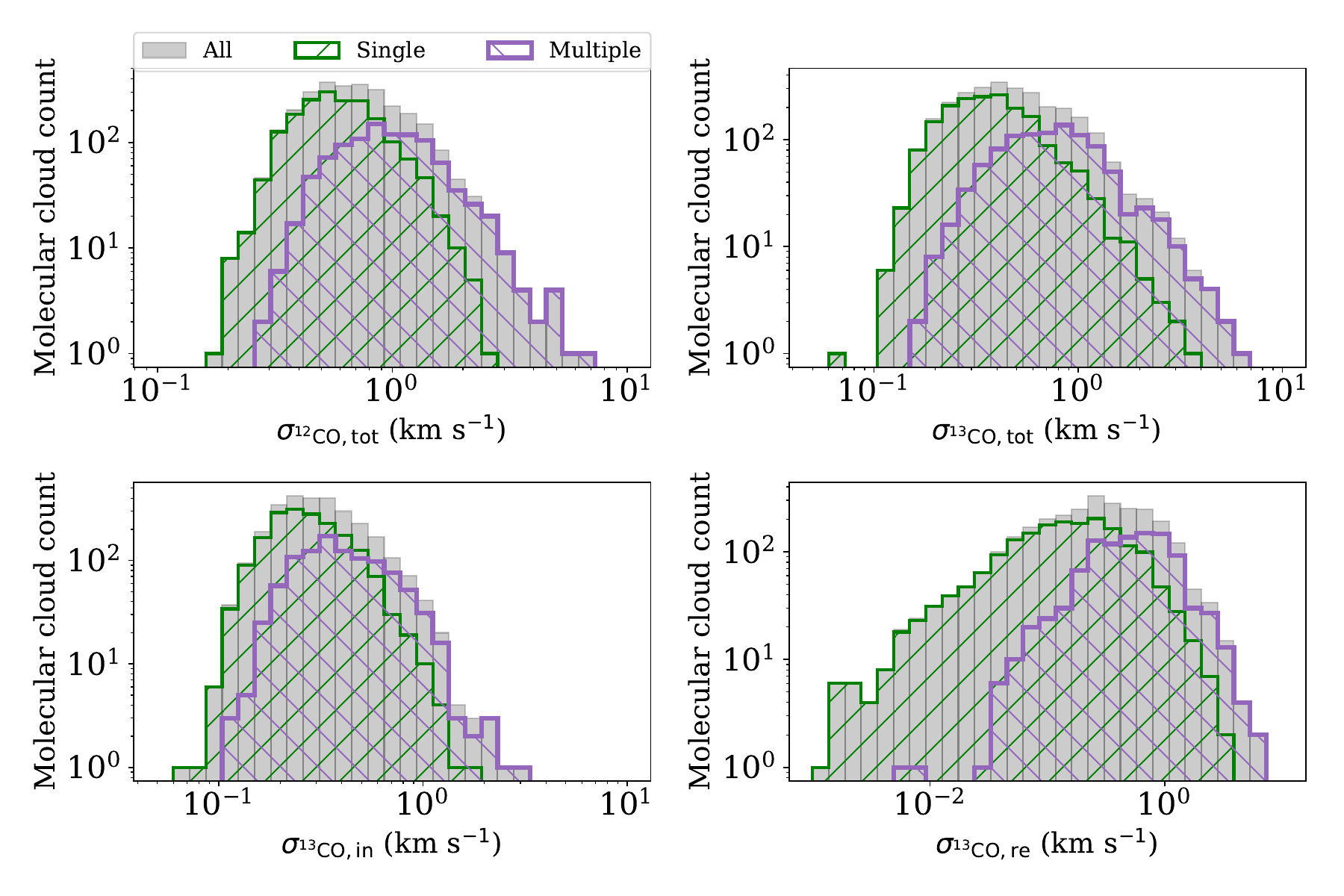} 
\caption{The distributions of the calculated values of $\sigma_{\rm ^{12}CO, tot}$, 
$\sigma_{\rm ^{13}CO,tot}$, $\sigma_{\rm ^{13}CO,in}$ and $\sigma_{\rm ^{13}CO,re}$ for MCs. 
Among that, the `All' represent the whole 2851 MCs, the `Single' and `Multiple' correspond to 
the MCs having single and multiple (more than one) $^{13}$CO structures, respectively.\label{fig:f_veldis}}
\end{figure*}

\begin{figure*}[ht]
    \plotone{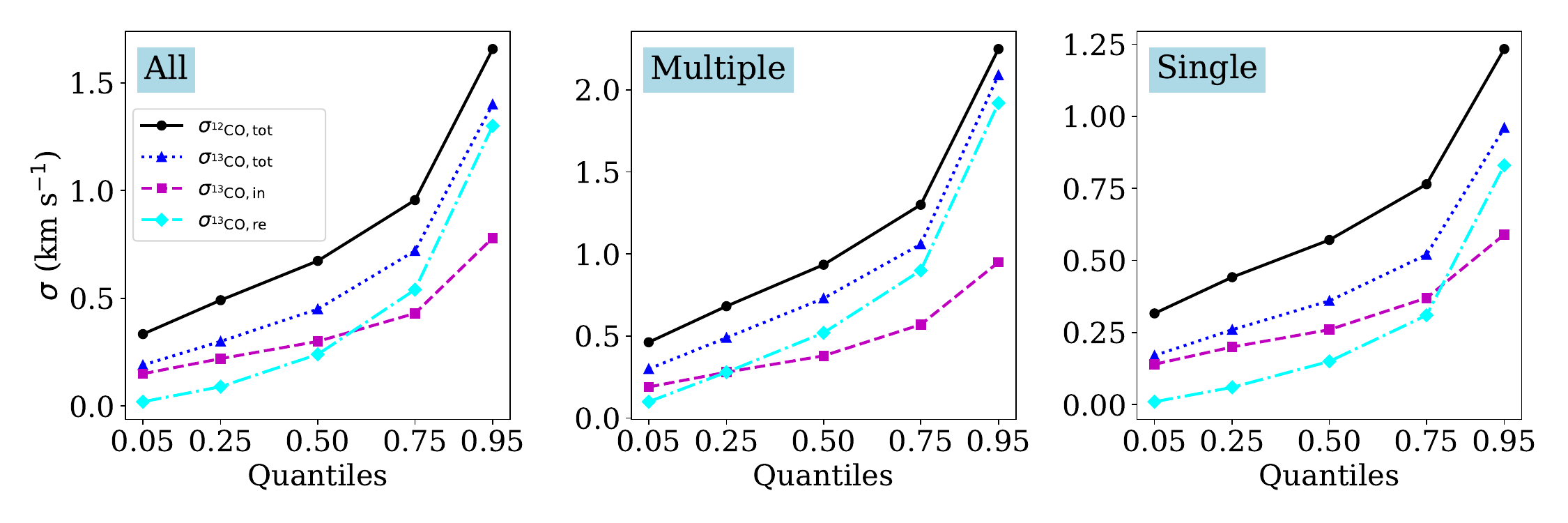}
    \caption{The increasing trends of the velocity dispersion components following the quantiles 
    at 0.05, 0.25, 0.5, 0.75, and 0.95 in their sequential data. 
    Among that, the `All' represent the whole 2851 MCs, the `Single' and `Multiple' correspond to 
the MCs having single and multiple (more than one) $^{13}$CO structures, respectively.\label{fig:f_velquan}}
\end{figure*}

Figure \ref{fig:f_veldis} shows the distributions of the calculated $\sigma_{\rm ^{12}CO,tot}$, $\sigma_{\rm ^{13}CO,tot}$,
$\sigma_{\rm ^{13}CO,in}$, and $\sigma_{\rm ^{13}CO,re}$ for each cloud. 
For comparison, these values for the MCs in the single and multiple regimes are also presented in Figure \ref{fig:f_veldis}. 
Moreover, their quantiles at 0.05, 0.25, 0.5, 0.75, 0.95 and the mean values are listed in Table \ref{tab:t_vel}. 
We find that the values of $\sigma_{\rm ^{12}CO,tot}$ and $\sigma_{\rm ^{13}CO,tot}$ in the `multiple' samples are 
systematically larger than those in the `single' regime. 
The $\sigma_{\rm ^{13}CO,tot}$ is composed of the $\sigma_{\rm ^{13}CO,re}$ and $\sigma_{\rm ^{13}CO,in}$.  
We find that the differences between the $\sigma_{\rm ^{13}CO,re}$ for the clouds in the `multiple' and the `single' 
are more obvious than those differences between the $\sigma_{\rm ^{13}CO,in}$ in the `single' and `multiple' regimes, as shown in Figure \ref{fig:f_veldis}.
From the values listed in Table \ref{tab:t_vel}, 
the $\sigma_{\rm ^{13}CO,re}$ in the `multiple' are larger than those in the `single' by a factor of $\sim$ 3, 
this factor is about 1.5 for the $\sigma_{\rm ^{13}CO,in}$.   
Figure \ref{fig:f_velquan} shows the increasing trends of these velocity dispersions at different 
quantiles as listed in Table \ref{tab:t_vel}. 
For the whole 2851 MCs, $\sim$ 65$\%$ of them are in the `single' regime and 
the rest $\sim$ 35$\%$ belong to the `multiple' regime. 
We find the increasing trend of $\sigma_{\rm ^{13}CO,tot}$ is similar to the $\sigma_{\rm ^{12}CO,tot}$, either in the multiple or single regime.
Nevertheless, the increasing trends of $\sigma_{\rm ^{13}CO,re}$ and $\sigma_{\rm ^{13}CO,in}$ are different, 
the $\sigma_{\rm ^{13}CO,re}$ increases with a steeper slope than that from the $\sigma_{\rm ^{13}CO,in}$. 
In the whole sample, we find $\sim$ 40$\%$ of them having the $\sigma_{\rm ^{13}CO,re}$ larger than the $\sigma_{\rm ^{13}CO,in}$, 
the fraction is $\sim$ 62$\%$ in the `multiple' samples and $\sim$ 20$\%$ in the `single' samples, respectively. 
From above results, those indicate the increase of $\sigma_{\rm ^{13}CO,tot}$ is mainly attributed to 
the increase of $\sigma_{\rm ^{13}CO,re}$ instead of the $\sigma_{\rm ^{13}CO,in}$, 
especially for the MC in the `multiple' regime. 

\begin{deluxetable*}{llcccccc}
\tablenum{2}
\tablecaption{Velocity dispersion components (km s$^{-1}$) \label{tab:t_vel}}
\tablewidth{0pt}
\tablehead{
\colhead{Velocity dispersion} & \colhead{Samples} & \colhead{0.05} & \colhead{0.25} &
\colhead{0.5} & \colhead{0.75} & \colhead{0.95} & \colhead{Mean} \\
}
\startdata
\multirow{3}{*}{$\sigma_{\rm ^{12}CO, tot}$} & All & 0.33 & 0.49 & 0.67 & 0.96 & 1.66 & 0.8 \\
                                                & Single & 0.32 & 0.44 & 0.57 & 0.76 & 1.23 & 0.64 \\
                                                & Multiple & 0.46 & 0.68 & 0.93 & 1.3 & 2.25 & 1.1 \\\hline
\multirow{3}{*}{$\sigma_{\rm ^{13}CO, tot}$} & All & 0.19 & 0.3 & 0.45 & 0.72 & 1.4 & 0.6 \\
                                                & Single & 0.17 & 0.26 & 0.36 & 0.52 & 0.96 & 0.44 \\
                                                & Multiple & 0.3 & 0.49 & 0.73 & 1.06 & 2.09 & 0.89 \\\hline
\multirow{3}{*}{$\sigma_{\rm ^{13}CO, re}$} & All & 0.02 & 0.09 & 0.24 & 0.54 & 1.30 & 0.41 \\ 
                                                & Single & 0.01 & 0.06 & 0.15 & 0.31 & 0.83 & 0.25 \\
                                                & Multiple & 0.1 & 0.28 & 0.52 & 0.90 & 1.92 & 0.70 \\\hline
\multirow{3}{*}{$\sigma_{\rm ^{13}CO, in}$} & All & 0.15 & 0.22 & 0.30 & 0.43 & 0.78 & 0.36 \\ 
                                                & Single & 0.14 & 0.20 & 0.26 & 0.37 & 0.59 & 0.30 \\
                                                & Multiple & 0.19 & 0.28 & 0.38 & 0.57 & 0.95 & 0.46 \\\hline
\enddata
\tablecomments{The quantiles at 0.05, 0.25, 0.5, 0.75, and 0.95 for each velocity dispersion (km s$^{-1}$) component in its sequential data and 
its mean value. Among that, the `All' represent the whole 2851 MCs, the `Single' and `Multiple' correspond to 
the MCs having single and multiple (more than one) $^{13}$CO structures, respectively.}
\end{deluxetable*}

\subsection{Correlations between kinetic energy compositions}
For a MC, the total velocity dispersion ($\sigma^{2}_{\rm ^{12}CO, tot}$) reflects the whole $^{12}$CO gas motion, 
the $\sigma^{2}_{\rm ^{13}CO, tot}$ is the velocity dispersion for the gas motion within its $^{13}$CO-bright region.
Meanwhile, the $\sigma^{2}_{\rm ^{13}CO, tot}$ is further decomposed into $\sigma^{2}_{\rm ^{13}CO, re}$ and $\sigma^{2}_{\rm ^{13}CO, in}$, 
corresponding to the internal and relative motions of $^{13}$CO structures, respectively. 
An interesting question concerns how these different components change with the increase of total velocity dispersion. 

\begin{figure*}[ht]
    \plotone{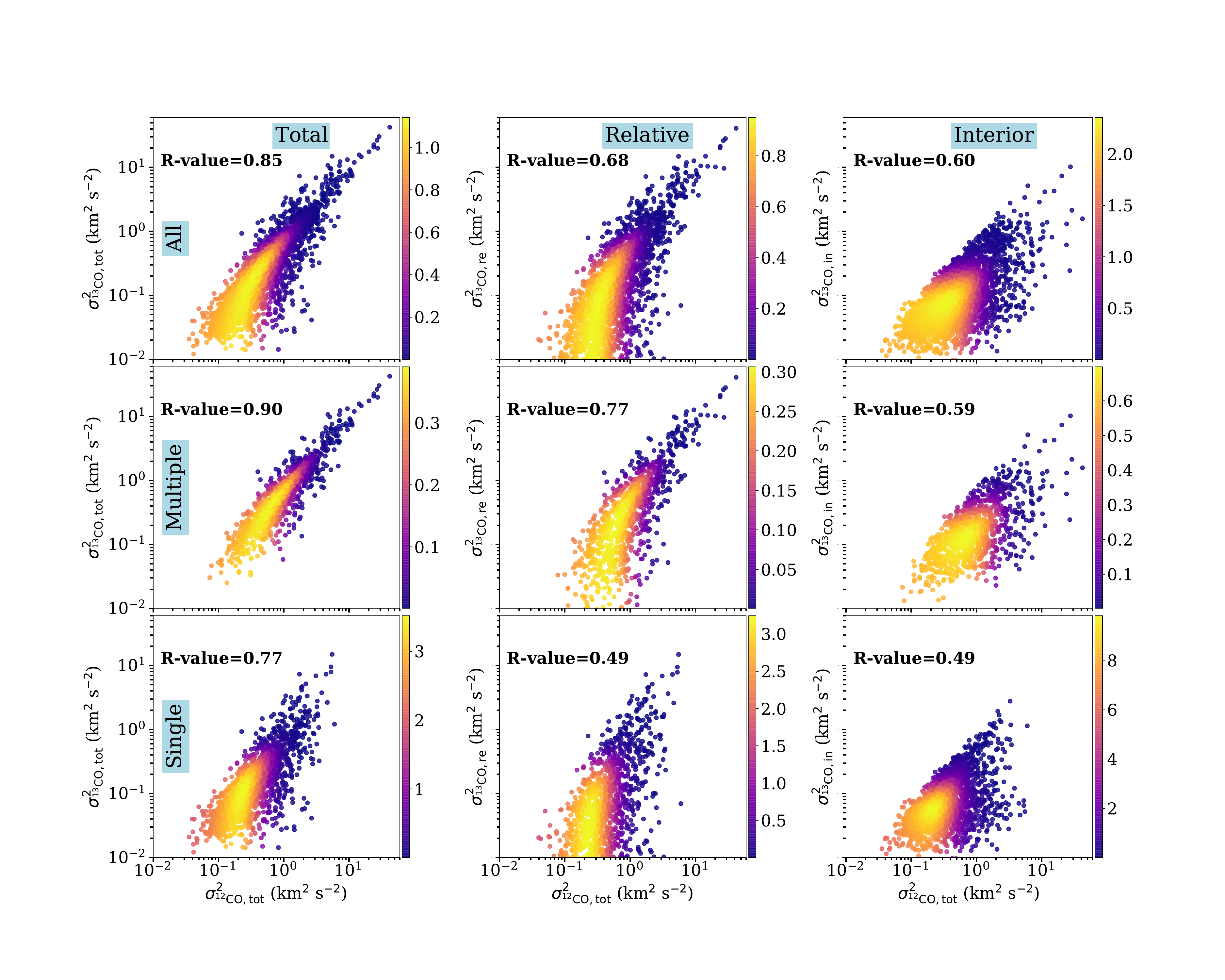} 
    \caption{Correlations between $\sigma^{2}_{\rm ^{13}CO,tot}$, $\sigma^{2}_{\rm ^{13}CO,re}$, 
    and $\sigma^{2}_{\rm ^{13}CO,in}$ with $\sigma^{2}_{\rm ^{12}CO,tot}$ for the MC samples in the 
    `all', `multiple', and `single' regimes, respectively. Each dot in the panels represents a $^{12}$CO MC. 
    The colors on these dots represent the distribution of the probability density function (2D-PDF) of 
    their $^{12}$CO MCs. The corresponding Spearman's rank correlation coefficient (R-value) is noted in each panel.\label{fig:f_corr1}}
\end{figure*}

In Figure \ref{fig:f_corr1}, we present the correlations between the velocity dispersions of $\sigma^{2}_{\rm ^{13}CO, tot}$, 
$\sigma^{2}_{\rm ^{13}CO,re}$ and $\sigma^{2}_{\rm ^{13}CO,in}$ with $\sigma^{2}_{\rm ^{12}CO,tot}$ in different MC samples, respectively. 
The Spearman's rank correlation coefficients (R-value) for these relations are also noted.
We find the $\sigma^{2}_{\rm ^{13}CO,tot}$ positively correlate with the $\sigma^{2}_{\rm ^{12}CO,tot}$ for all the samples 
with a R-value of 0.85. Meanwhile, this R-value of 0.9 for the MCs in the `multiple' regime is larger than the R-value of 0.77 in 
the `single' regime. The $\sigma^{2}_{\rm ^{13}CO,tot}$ is further decomposed into the $\sigma^{2}_{\rm ^{13}CO,re}$ and $\sigma^{2}_{\rm ^{13}CO,in}$.   
The R-value is 0.68, 0.77, and 0.49 for the relations between $\sigma^{2}_{\rm ^{13}CO,re}$ and $\sigma^{2}_{\rm ^{12}CO,tot}$ 
for the MC samples in the `all', `multiple', and `single', respectively. 
The corresponding R-value is 0.60, 0.59, and 0.49 for these relations between $\sigma^{2}_{\rm ^{13}CO,in}$ and $\sigma^{2}_{\rm ^{12}CO, tot}$. 
We find that the $\sigma^{2}_{\rm ^{13}CO,re}$ are more positively correlated with $\sigma^{2}_{\rm ^{12}CO,tot}$ in the `multiple' regime, 
while the $\sigma^{2}_{\rm ^{13}CO,in}$ don't significantly increase with $\sigma^{2}_{\rm ^{12}CO,tot}$, 
either in the `single' or `multiple' regime. 
That further indicates the increase of $\sigma^{2}_{\rm ^{12}CO,tot}$ is mainly attributed to the increase of $\sigma^{2}_{\rm ^{13}CO,re}$, 
i.e. the relative motion between $^{13}$CO gas structures, especially for the MCs in the `multiple' regime. 

\begin{figure*}[ht]
    \plotone{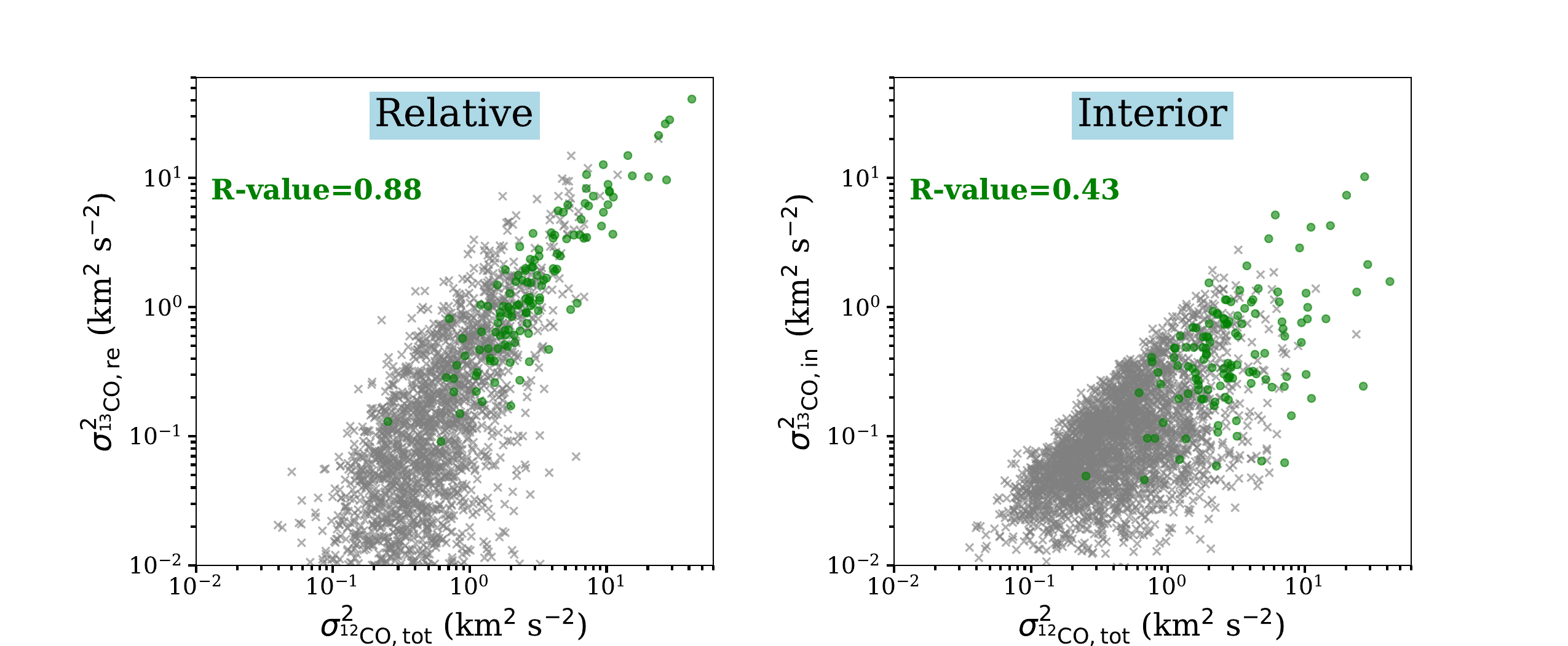} 
    \caption{Correlations between $\sigma^{2}_{\rm ^{13}CO,re}$ with $\sigma^{2}_{\rm ^{12}CO,tot}$ (\textbf{left panel})
    and $\sigma^{2}_{\rm ^{13}CO,in}$ with $\sigma^{2}_{\rm ^{12}CO,tot}$ (\textbf{right panel}) for the MC samples, respectively. 
    The green dots in the panels represent $^{12}$CO MCs harboring at least ten $^{13}$CO structures, 
    the corresponding Spearman's rank correlation coefficient (R-value) is noted in each panel. 
    The gray crosses are for the left $^{12}$CO MCs, whose number of $^{13}$CO structures is less than 10.\label{fig:f_corr2}}
\end{figure*}

We should note that the statistical error of the $\sigma_{\rm ^{13}CO,re}$ is 
influenced by the number of $^{13}$CO structures within the MC. In our 2851 MC samples, 
$\sim$ 65$\%$ of clouds harbor a single $^{13}$CO structure, about 15$\%$ of them have double 
$^{13}$CO structures, and the rest $\sim$ $20\%$ have at least three $^{13}$CO structures, as presented in our Paper II.
Thus the calculated $\sigma^{2}_{\rm ^{13}CO,re}$ of MCs are scattered in statistics 
and it should be reliable for MCs having at least ten $^{13}$CO structures.  
In Figure \ref{fig:f_corr2}, we highlight the 125 clouds having at least ten $^{13}$CO structures. 
For these 125 MCs, the correlation between their $\sigma^{2}_{\rm ^{13}CO,re}$ and $\sigma^{2}_{\rm ^{12}CO,tot}$ 
has a higher R-value of 0.88, however, the R-value is 0.43 for the relation between their
$\sigma^{2}_{\rm ^{13}CO,in}$ and $\sigma^{2}_{\rm ^{12}CO,tot}$. 
That further demonstrates the relative motion between $^{13}$CO structures within clouds, 
instead of their interior gas motion, is more positively correlated with their global velocity dispersions.  

\begin{figure*}[ht]
    \plotone{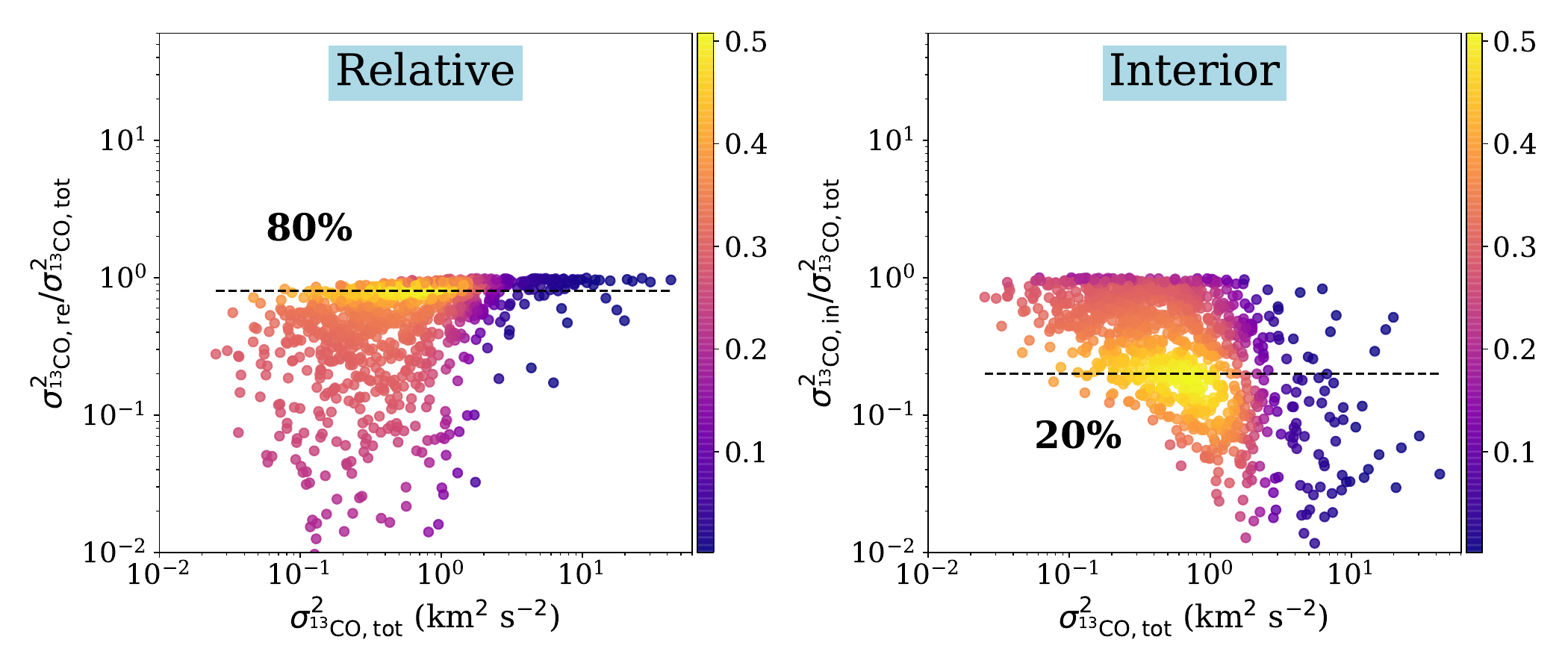} 
    \caption{Left panel: Relation between the ratio of $\sigma^{2}_{\rm ^{13}CO,re}$ over $\sigma^{2}_{\rm ^{13}CO,tot}$ 
    with the $\sigma^{2}_{\rm ^{13}CO,tot}$. Right panel: Relation between the ratio of $\sigma^{2}_{\rm ^{13}CO,in}$ over $\sigma^{2}_{\rm ^{13}CO,tot}$ 
    with the $\sigma^{2}_{\rm ^{13}CO,tot}$. Each dot in both panels represents a MC in the `multiple' regime.  
    The colors on the dots represent the distribution of the probability density function (2D-PDF) of 
    their $^{12}$CO MCs. The black-dashed lines show the values of $\sigma^{2}_{\rm ^{13}CO,re}$/$\sigma^{2}_{\rm ^{13}CO,tot}$ = 0.8 
    in the left panel and $\sigma^{2}_{\rm ^{13}CO,in}$/$\sigma^{2}_{\rm ^{13}CO,tot}$ = 0.2 in the right panel. \label{fig:f_frac}}
\end{figure*}

Since the two types of gas motions from the $^{13}$CO structures, i.e. their relative motions ($\sigma^{2}_{\rm ^{13}CO,re}$) 
and internal motions ($\sigma^{2}_{\rm ^{13}CO,in}$), exhibit the different correlations with $\sigma^{2}_{\rm ^{12}CO,tot}$, 
especially for the MC samples in the `Multiple' regime. 
We further look into the fractions of the $\sigma^{2}_{\rm ^{13}CO,re}$ and $\sigma^{2}_{\rm ^{13}CO,in}$ within $\sigma^{2}_{\rm tot, ^{13}CO}$,
and reveal how the fractions change with the increases of $\sigma^{2}_{\rm ^{13}CO,tot}$. 
Figure \ref{fig:f_frac} shows the relations between $\sigma^{2}_{\rm ^{13}CO,re}$/$\sigma^{2}_{\rm ^{13}CO,tot}$ and $\sigma^{2}_{\rm ^{13}CO,tot}$ 
for the MCs in the `multiple' regime, as well as the relations between $\sigma^{2}_{\rm ^{13}CO,in}$/$\sigma^{2}_{\rm ^{13}CO,tot}$ and $\sigma^{2}_{\rm ^{13}CO,tot}$. 
We find that the fractions of $\sigma^{2}_{\rm ^{13}CO,re}$ within $\sigma^{2}_{\rm ^{13}CO,tot}$ tend to be 
$\sim$ 80 percent, which also has a slight increase as the increase of $\sigma^{2}_{\rm ^{13}CO,tot}$, 
the left $\sim$ 20 percent contributions are from the $\sigma^{2}_{\rm ^{13}CO,in}$. 
This means the relative motions of $^{13}$CO gas structures tend to be the dominant form to store the kinetic energy for MCs in the `multiple' regime.

\section{Discussion}

\subsection{Centroid velocities of $^{13}$CO line emission as the systematic velocities of MCs}
In order to ensure the effects of the V$_{\rm cen, ^{12}CO}$ or V$_{\rm cen, ^{13}CO}$ as systematic velocity on our results, 
here, we take the V$_{\rm cen, ^{13}CO}$ as the systematic velocities of MCs and further calculate the velocity dispersions following Eq.\ref{e:decom}. 
For the MCs in the `single' regime, the systematic velocities(V$_{\rm cen, ^{13}CO}$) of MCs are 
consistent with the centroid velocities of their single $^{13}$CO structures. 
Thus the relative velocity between the $^{13}$CO structure and systematic velocity for a MC in the `single' regime 
is equal to zero. 
The MCs in the `All' regime can be divided into MCs in the `Multiple' and `Single' regimes.
In Figure \ref{fig:f_corr2_13cen}, we show the relations between $\sigma^{2}_{\rm ^{12}CO, tot}$ and 
$\sigma^{2}_{\rm ^{13}CO, tot}$ for the MCs, which are in the `All', `Multiple', and `Single', respectively. 
The R-values are 0.81 for the MCs in the `Multiple' and 0.41 for those in the `Single'. 
That implies that the $\sigma^{2}_{\rm ^{13}CO, tot}$ tend to increase with $\sigma^{2}_{\rm ^{12}CO, tot}$ for MCs 
in the `Multiple' regime. However, the $\sigma^{2}_{\rm ^{13}CO, tot}$ for the MC in the `Single' regime, 
which is equal to its $\sigma^{2}_{\rm ^{13}CO, in}$, doesn't significantly increase with $\sigma^{2}_{\rm ^{12}CO, tot}$.  

In addition, the $\sigma^{2}_{\rm ^{13}CO, tot}$ is decomposed into $\sigma^{2}_{\rm ^{13}CO, re}$ and $\sigma^{2}_{\rm ^{13}CO, in}$. 
Figure \ref{fig:f_corr1_13cen} presents the correlations between 
the $\sigma^{2}_{\rm ^{13}CO, tot}$, $\sigma^{2}_{\rm ^{13}CO, re}$, $\sigma^{2}_{\rm ^{13}CO, in}$ with 
the $\sigma^{2}_{\rm ^{12}CO, tot}$ for all the samples, respectively. 
We find that the $\sigma^{2}_{\rm ^{13}CO, re}$ tends to increase with $\sigma^{2}_{\rm ^{12}CO, tot}$ having a R-value of 0.71, 
while the $\sigma^{2}_{\rm ^{13}CO, in}$ doesn't have a clear trend with $\sigma^{2}_{\rm ^{12}CO, tot}$, 
whose R-value is 0.54.  

Furthermore, we decompose the $\sigma^{2}_{\rm ^{13}CO, tot}$ for the MCs in the `Multiple' regime. 
Figure \ref{fig:f_corr3_13cen} presents the correlations between the $\sigma^{2}_{\rm ^{13}CO, tot}$, 
$\sigma^{2}_{\rm ^{13}CO, re}$, $\sigma^{2}_{\rm ^{13}CO, in}$ with the $\sigma^{2}_{\rm ^{12}CO, tot}$ for the samples in the `Multiple', respectively.
For the MCs in the `Multiple' regime, the $\sigma^{2}_{\rm ^{13}CO, tot}$ also tends to 
increase with $\sigma^{2}_{\rm ^{12}CO, tot}$, whose R-value is 0.81. 
Meanwhile, the R-value is 0.71 for the relation between $\sigma^{2}_{\rm ^{13}CO, re}$ and $\sigma^{2}_{\rm ^{12}CO, tot}$ 
and 0.54 for the relation between $\sigma^{2}_{\rm ^{13}CO, in}$ and $\sigma^{2}_{\rm ^{12}CO, tot}$. 
The relations between $\sigma^{2}_{\rm ^{13}CO, re}$ and $\sigma^{2}_{\rm ^{12}CO, tot}$ are 
consistent either in the `all' or `multiple' regime, 
due to that the $\sigma^{2}_{\rm ^{13}CO, re}$ is zero for in MCs in the `single' regime. 
In addition, the relations between $\sigma^{2}_{\rm ^{13}CO, in}$ and $\sigma^{2}_{\rm ^{12}CO, tot}$ have 
the same R-value either in the `all' or `multiple'. 
The $\sigma^{2}_{\rm ^{13}CO, re}$ is always more positively correlated with the $\sigma^{2}_{\rm ^{12}CO, tot}$ 
than $\sigma^{2}_{\rm ^{13}CO, in}$ for the MCs in the `multiple' regime.

Thus, either V$_{\rm cen, ^{12}CO}$ or V$_{\rm cen, ^{13}CO}$ is defined as the systematic velocity for a MC, 
we find the relative motion between $^{13}$CO structures gradually provides the primary contributions to the $\sigma^{2}_{\rm ^{12}CO, tot}$ 
with the development of MC scales.

\subsection{Effects of optical depths}
Since the opacity can severely affect the observed linewidths in the optically thick CO line emission, e.g. $^{12}$CO(1-0) lines \citep{Phillips1979, Hacar2016, Pineda2008}. 
It is necessary to explore and constrain the influence of the opacity broadening on the derived $\sigma_{\rm ^{12}CO, tot}$. 
The spectral profiles of $^{12}$CO and $^{13}$CO lines emission are described by the radiative transfer equation 
\citep{Rohlfs1996}:

\begin{equation}\label{e:tau12}
T_{\rm mb, ^{12}CO}=(J_\nu(T_{\rm ex})- J_{\nu}(T_{\rm bg}))(1-e^{-\tau_{\rm ^{12}CO}}) ,
\end{equation}
\begin{equation}\label{e:tau13}
T_{\rm mb, ^{13}CO}=(J_\nu(T_{\rm ex})- J_{\nu}(T_{\rm bg}))(1-e^{-\tau_{\rm ^{13}CO}}) ,
\end{equation}
where $J_{\nu}(T)=(\frac{h\nu/k}{exp(h\nu/kT)-1})$, $T_{\rm ex}$ is the excitation temperature of the lines,
$T_{\rm bg}$= 2.7 K is the cosmic microwave background temperature, 
$\tau_{\rm ^{12}CO}$ and $\tau_{\rm ^{13}CO}$ are the line opacities at the $^{12}$CO and $^{13}$CO emission, respectively. 
In the $^{13}$CO-bright region, 
we assume that the excitation temperature of $^{13}$CO(1-0) line is the same as that for the $^{12}$CO(1-0) line, 
then combine the Eq. \ref{e:tau12} and Eq. \ref{e:tau13} as: 

\begin{equation}\label{e:tau_13bri}
    \frac{T_{\rm mb, ^{12}CO}}{T_{\rm mb, ^{13}CO}} \approx \frac{1-e^{-\tau_{\rm ^{12}CO}}}{1-e^{-\tau_{\rm ^{13}CO}}} ,
\end{equation}

Considering the typical relative abundances for the $^{12}$CO and $^{13}$CO isotopologues in the local ISM, 
i.e., X($^{13}$CO):X($^{12}$CO)=7.3:560 in \cite{Wilson1994}, 
the opacities of the $^{12}$CO(1-0) and $^{13}$CO(1-0) line emission can be related by their relative 
abundances as $\tau_{\rm ^{12}CO} \simeq X(\rm ^{12}CO)/X(\rm ^{13}CO) \times \tau_{\rm ^{13}CO}$.

To visualize the distribution of the $^{12}$CO and $^{13}$CO line emission in a MC, 
In Figure \ref{fig:ftau_trend1} and \ref{fig:ftau_trend2}, 
we present the distributions of the velocity-integrated intensities of $^{12}$CO (I$_{\rm ^{12}CO}$) and $^{13}$CO (I$_{\rm ^{13}CO}$) emission for two MC samples 
and also the distributions of the pixel numbers in the intervals of the values (I$_{\rm ^{12}CO}$, I$_{\rm ^{13}CO}$, $\tau_{\rm ^{13}CO}$).  
We find that the $^{12}$CO line intensities in the pixels within the $^{13}$CO-dark region vary smoothly, 
especially comparing with those values within the $^{13}$CO-bright region.  
Thus it is reasonable to determine the excitation temperature in the periphery of $^{13}$CO-bright region (T$_{0}$) 
to be consistent with that in the $^{13}$CO-dark region.
Further the optical opacity of $^{12}$CO emission in the $^{13}$CO-dark region can be estimated as:
\begin{equation}\label{e:tau_13dark}
\frac{T_{\rm mb, ^{12}CO}}{T_{\rm mb, ^{12}CO,0}} \approx \frac{1-e^{-\tau_{\rm ^{12}CO}}}{1-e^{-\tau_{\rm ^{12}CO, 0}}} ,
\end{equation}
where $T_{\rm mb, ^{12}CO, 0}$ and $\tau_{\rm ^{12}CO, 0}$ represent the brightness temperature and 
the optical depth of $^{12}$CO line emission lie at the periphery of the $^{13}$CO-bright region. 
We adopt the minimum value of $\tau_{\rm ^{12}CO}$ within the $^{13}$CO-bright region 
as the $\tau_{\rm ^{12}CO, 0}$, and its corresponding $T_{\rm mb, ^{12}CO}$ is defined as $T_{\rm mb, ^{12}CO, 0}$.

\begin{figure*}[ht]
    \plotone{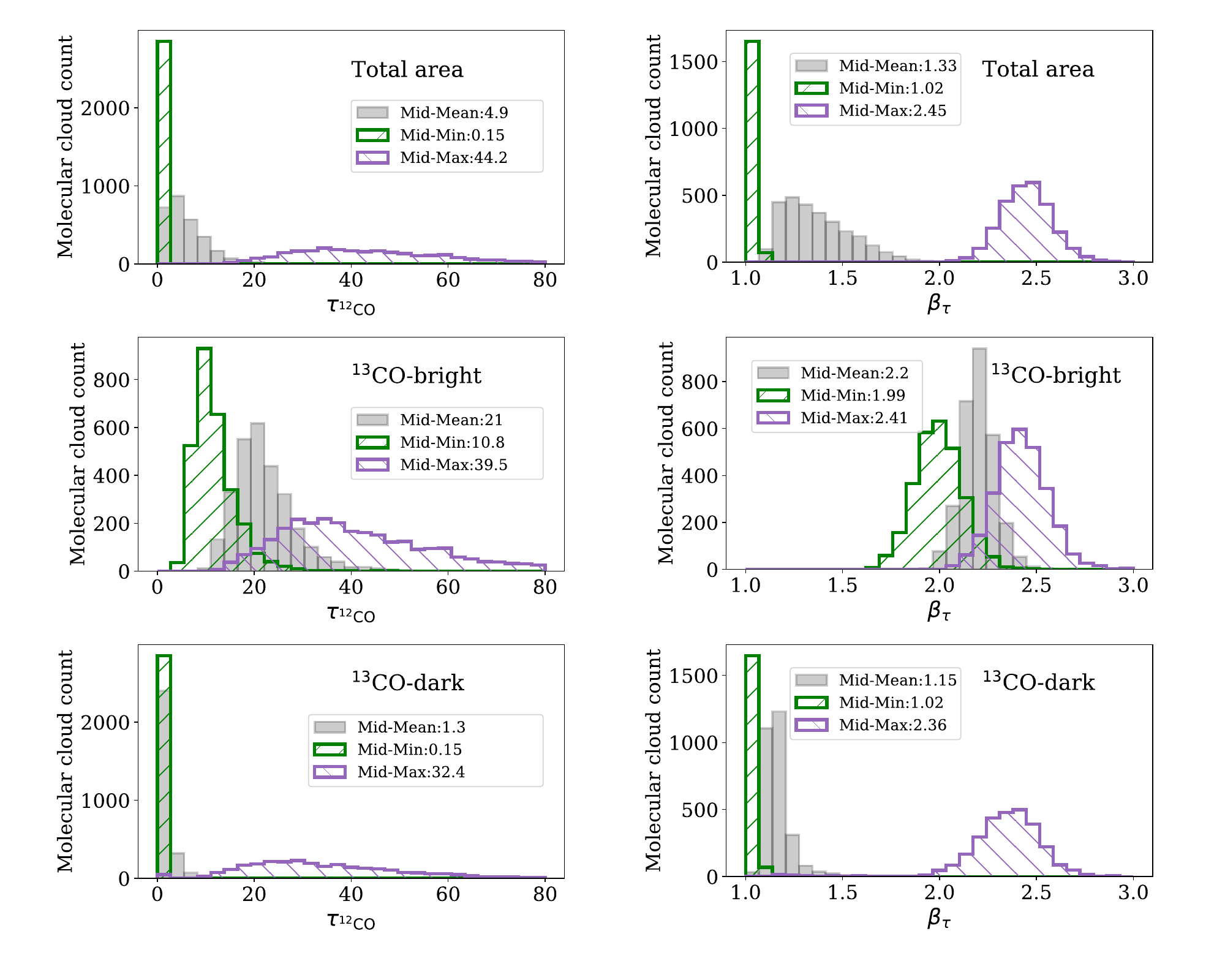} 
    \caption{Distributions of the optical depths of $^{12}$CO emission ($\tau_{\rm ^{12}CO}$) and $\beta_{\tau}$ in the whole regions, $^{13}$CO-bright regions, 
    and $^{13}$CO-dark regions of our 2851 MC samples, respectively. 
    The green histograms represent the distributions of the minimal values of $\tau_{\rm ^{12}CO}$ and $\beta_{\tau}$ in the 
    whole regions, $^{13}$CO-bright regions, and $^{13}$CO-dark regions of MCs, respectively. 
    The gray histograms are for their mean values and the magenta histograms are for their maximum values. 
    The median values for the distributions of these minimal, mean, and maximum $\tau_{\rm ^{12}CO}$ and $\beta_{\tau}$ 
    in the MC samples are noted in each panel. \label{fig:ftau_dist}}
\end{figure*}

Thus using Eq.\ref{e:tau_13bri} and Eq.\ref{e:tau_13dark}, 
we drived the opacities of $^{12}$CO line emission in the $^{13}$CO-bright and $^{13}$CO-dark region of each MC, respectively. 
Figure \ref{fig:ftau_dist} shows the distributions of the minimal ($\tau_{\rm ^{12}CO, min}$), maximal ($\tau_{\rm ^{12}CO, max}$), and mean opacities ($\bar\tau_{\rm ^{12}CO}$) 
of the $^{12}$CO emission in the whole region in each MC, as well as in its $^{13}$CO-dark and $^{13}$CO-bright regions, respectively.
The medians for these values are also noted in each panel. 
The $\bar\tau_{\rm ^{12}CO}$ in the whole regions of MCs have a median value of 4.9, 
this value is 21 in the $^{13}$CO-bright regions of MCs and 1.3 in the $^{13}$CO-dark regions. 
That indicates the $^{12}$CO emission in the $^{13}$CO-bright region is optically thick, 
while that in the $^{13}$CO-dark region is slightly optically-thick. 
From the $\tau_{\rm ^{12}CO, max}$ distribution, 
we find that $\sim$ 90$\%$ of our samples with $\tau_{\rm ^{12}CO, max}$ less than 80. 
That indicates the most of $^{13}$CO emission(X($^{13}$CO):X($^{12}$CO)=1:76.7) is optically thin in these MCs. 
In addition, the median of $\tau_{\rm ^{12}CO, min}$ in the $^{13}$CO-bright regions of our MC samples is 10.8, 
which may be the critical value between the $^{13}$CO-bright and $^{13}$CO-dark regions, 
under our observational sensitivities of 0.25 K at a velocity resolution of $\sim$ 0.2 km s$^{-1}$ for the $^{13}$CO line emission. 

Since the $^{12}$CO line emission in the $^{13}$CO-bright region is optically thick, 
the line broadening produced by the line opacity can be defined as \citep{Phillips1979}:
\begin{equation}\label{e:beta}
\beta_{\tau}=\frac{\Delta V}{\Delta V_{\rm int}} = \frac{1}{\sqrt{ln2}}\left[ln\left(\frac{\tau}{ln\left(\frac{2}{e^{-\tau+1}}\right)}\right)\right]^{1/2} ,   
\end{equation}
where $\Delta V$ and $\Delta V_{\rm int}$ are referred to as the observed and the intrinsic velocity width (FWHM), respectively. 
We use the $\tau_{\rm ^{12}CO}$ map to estimate the $\beta$ map for each cloud according to Eq.\ref{e:beta}.  
Figure \ref{fig:ftau_dist} shows the distributions of the minimal ($\beta_{min}$), maximal ($\beta_{max}$), and mean $\beta$ ($\bar\beta$) values 
in the whole region of each MC, as well as in the $^{13}$CO-dark and $^{13}$CO-bright regions, respectively. 
The $\bar\beta$ in the whole regions of MCs range from $\sim$ 1 to 2.1 with a median of 1.33, 
the median value is 2.2 and 1.15 for the $\bar\beta$ in the $^{13}$CO-bright regions and the $^{13}$CO-dark regions of MCs, respectively. 

We use the value of $\bar\beta$ in the whole region of each cloud to mitigate the effect of $^{12}$CO line opacity on the 
velocity dispersion of $^{12}$CO emission, following $\sigma_{\rm ^{12}CO,int}$ = $\sigma_{\rm ^{12}CO,tot}/\bar\beta$. 
Figure \ref{fig:ftau_rev} shows the relations between the $\sigma^{2}_{\rm ^{13}CO,tot}$, 
$\sigma^{2}_{\rm ^{13}CO,re}$, $\sigma^{2}_{\rm ^{13}CO,in}$ with the $\sigma^{2}_{\rm ^{12}CO,int}$. 
We find that the Spearman's rank correlation coefficients (R-value) in the relations 
between $\sigma^{2}_{\rm ^{13}CO,re}$ and $\sigma^{2}_{\rm ^{12}CO,int}$ are higher than 
those for the relations between $\sigma^{2}_{\rm ^{13}CO,re}$ and $\sigma^{2}_{\rm ^{12}CO,tot}$. 
However, the R-value for the relations between $\sigma^{2}_{\rm ^{13}CO,in}$ and $\sigma^{2}_{\rm ^{12}CO,int}$ 
are lower than those from the relations between $\sigma^{2}_{\rm ^{13}CO,in}$ and $\sigma^{2}_{\rm ^{12}CO,tot}$. 
That means our conclusion that the relative motions between $^{13}$CO structures mainly account for 
the increases of $\sigma^{2}_{\rm ^{12}CO,int}$ in the MCs with `multiple' $^{13}$CO structures is still established, 
after taking the opacity effects into account. 

\subsection{Kinetic energy in the molecular cloud: microscopic versus macroscopic} 

Larson's relation between MCs, a power-law relation between their total velocity dispersions and spatial sizes, 
is similar to the Kolmogorov law for incompressible turbulence \citep{Larson1981}. 
This suggests the interstellar turbulent energy cascade, 
i.e. the energy range of random motion produced on a large spatial scale, 
then cascades to smaller scales in the inertial range and 
finally dissipates in the damping range \citep{Baker1976, Fleck1980}. 
However, the MCs are quite inhomogeneous and highly structured, consisting of numerous denser clumps and cores, 
Larson's relations are thought to be violated in these dense regions within MCs \citep{Heyer2009, Ballesteros2011, Traficante2018}.

Our main observational result is that the relative motions between distinct $^{13}$CO structures gradually dominate 
the total velocity dispersions of MCs with the development of MC scales. 
The random motion between discrete internal structures is a manifestation of macroscopic turbulence (macro-turbulence),  
which is thought to primarily determine the line widths of spectral profiles \citep{Zuckerman1974, Silk1985, Kwan1986, Hacar2016}.
The existence of dense structures and bulk kinetic energy on various scales 
provide evidence for the compressibility of MCs.   
Such a model also would allow a fairly long time scale for damping the motions of internal structures 
and further tend to prevent the entire cloud from collapsing rapidly \citep{Zuckerman1974}. 
The macro-turbulent could be generated by either large-scale or small-scale converging flows \citep{Klessen2000}. 
Taking the complex Galactic environment into consideration, 
the gas dynamics within MCs manifested as the macro-turbulence 
could be driven by the hierarchical gravitational collapse \citep{Zuckerman1974, Baker1976, Ballesteros2011}, 
stellar feedbacks, such as stellar wind and outflows \citep{Silk1985}, supernova explosions \citep{Watkins2023, Skarbinski2023}, and spiral shocks \citep{Bonnell2006, Falceta2015}.

\subsection{Implications on the dynamic evolution of molecular clouds}
In our series of works using a large sample of $^{12}$CO MCs, 
we have investigated these MCs in different aspects, 
including their morphologies (Paper I), dense gas fractions traced by $^{13}$CO lines (Paper II), 
spatial distributions of their internal $^{13}$CO structures (Paper III), 
and internal kinetic energy depositions in this work. 
The morphologies of $^{12}$CO MCs are developing from non-filaments (e.g. clumpy and extended structures) to filaments 
with the increases of their scales. 
After extracting the relatively dense gas structures traced by $^{13}$CO lines within each $^{12}$CO cloud, 
we find that the $^{13}$CO gas contents within $^{12}$CO MCs are confined by the scales of $^{12}$CO MCs (Paper II). 
Furthermore, we also reveal that there is a preferred spatial separation between $^{13}$CO structures within these $^{12}$CO MCs, 
independent of their spatial scales. 
In this work, we further find that the relative motions of $^{13}$CO structures gradually provide the primary contributions to 
the total velocity dispersions of MCs, as the development of MC scales.
Combining these observational results, we find that MCs tend to exhibit complex and filamentary morphology, 
local density enhancements ($^{13}$CO gas structures), structural stabilization (regularly-spaced $^{13}$CO gas structures), 
and relative motions between $^{13}$CO structures, along with the development of MC scales. 

Taking the dynamic environment of the Galaxy into account, 
e.g. the converging gas flows driven by galactic differential 
rotation and shear, large-scale instabilities, and stellar feedback from the supernova explosions and HII regions. 
Some numerical simulations predict that converging flows 
could induce shock compressions and produce local density enhancements in the diffuse ISM \citep{Scalo1998,Vazquez1995,Ballesteros1999, MacLow2004, McKee2007,Ballesteros2020}, 
and also are the primary drivers of cloud mergers and splits \citep{Tasker2009, Dobbs2012, Dobbs2013, Dobbs2015, Skarbinski2023, Watkins2023}. 
We propose an alternative picture for the assembly and destruction of MCs: the regularly-spaced $^{13}$CO structures 
are thought to be the building bricks of MCs, the dynamic build-up and destruction of MCs proceeds under this fundamental brick. 

We further compare this picture with the simulated results in terms of the 
development of the velocity structures, density structures, and morphology of MCs. 
Firstly, we compare the relative velocities in the simulated mergers and 
the relative velocities between $^{13}$CO structures within MCs. 
In our MC samples, we find $\sim$ 90$\%$ of the radially relative velocity between 
internal $^{13}$CO structures is less than 5 km s$^{-1}$, as presented in Paper III. 
Also, 95$\%$ of MCs have velocity dispersions of $\sigma_{\rm ^{12}CO,tot}$ less than 1.66 km s$^{-1}$, as listed in Table 2. 
These observational results are consistent with that mergers are typically slow, 
occurring at relative speeds of $\lesssim$ 5 km/s 
($\sim$ 3 times the internal cloud velocity dispersion) predicted in the simulation of \cite{Dobbs2015,Skarbinski2023}, 
which is thought to be unlikely to cause the shocks \citep{Balfour2015, Balfour2017, Liow2020}. 
In addition, the relative motion between $^{13}$CO structures provides the main contribution to 
the increases in the global velocity dispersion for MCs, 
this is also consistent with the cloud mergers leading to the higher velocity dispersions of MCs, 
as simulated in \cite{Dobbs2011, Dobbs2015, Jeffreson2021, Skarbinski2023}.

Secondly, the slow merger also is consistent with the existence of a preferred 
separation between $^{13}$CO structures within MCs, independent of the MC scales. 
The mergers are typically slow and gentle so that it is unable to induce the shocks 
and to further change the distributions of their density structures \citep{Balfour2015, Balfour2017, Liow2020}. 
That is also consistent with the numerical results of slow mergers that do not have a strong impact on the 
density structures of clouds in the works of \cite{Dobbs2015, Jeffreson2021}.

Lastly, \cite{Dobbs2015} discussed that the 
merger of clouds tends to result in an even further elongated cloud, i.e. 
the smaller clouds gently merge onto the ends of the larger clouds, 
which tend to align with spiral arms and interact along their minor axis. 
This process is also coincident with that clouds tend to exhibit from non-filaments to filaments as increasing with cloud scales \citep{Yuan2021}.

According to the comparisons between our observational facts with the simulated results, 
we suggest the assembly and destruction of MCs: the regularly-spaced $^{13}$CO structures 
are thought to be the building bricks of MCs, and the dynamic build-up of MCs 
proceed by slow mergers among these fundamental bricks, this process does not suffer a significant change on 
their density structures, but have an impact on their global velocity structures.

\section{Conclusions}
Using a sample of 2851 $^{12}$CO molecular clouds, inside which a total of 9566 $^{13}$CO gas structures are identified, 
we investigate the relations between the internal and relative gas motions of $^{13}$CO structures with 
the total velocity dispersions of each $^{12}$CO cloud, respectively. Our main conclusions are as follows: 

1. The centroid velocities calculated by $^{12}$CO and $^{13}$CO lines emission are nearly consistent, 
their differences are less than $\sim$ 0.65 km s$^{-1}$ in 90$\%$ of the whole MC samples and less than $\sim$ 0.15 km s$^{-1}$ in 
50$\%$ of the samples.

2. The increasing trend of $\sigma_{\rm ^{13}CO,tot}$ is similar to that of $\sigma_{\rm ^{12}CO,tot}$. 
For its components of $\sigma_{\rm ^{13}CO,re}$ and $\sigma_{\rm ^{13}CO,in}$,  
the $\sigma_{\rm ^{13}CO,re}$ increases with a slope, which is steeper than that from $\sigma_{\rm ^{13}CO,in}$. 

3. The relation between $\sigma^{2}_{\rm ^{13}CO,re}$ and $\sigma^{2}_{\rm ^{12}CO,tot}$ is more positively correlated than 
that between $\sigma^{2}_{\rm ^{13}CO,in}$ and $\sigma^{2}_{\rm ^{12}CO,tot}$. 
This provides a clear trend of macro-turbulence becoming the dominant component of kinetic energy with the development of MC scales. 

4. Comparing our observational results with the simulations, 
we propose a picture on the assembly and destruction of MCs: the regularly-spaced $^{13}$CO structures 
are thought to be the building bricks of MCs, and the transient processes of MCs 
proceed by slow mergers among these fundamental bricks, 
during which the density structures of MCs do not vary significantly, 
but their global velocity structures are influenced.

\begin{acknowledgments}
    This research made use of the data from the Milky Way Imaging Scroll Painting (MWISP) project, 
    which is a multi-line survey in $^{12}$CO/$^{13}$CO/C$^{18}$O along the northern galactic plane with PMO-13.7m telescope. 
    We are grateful to all of the members of the MWISP working group, particulaly the staff members at the PMO-13.7m telescope, 
    for their long-term support. This work was supported by the National Natural Science Foundation of China through grant 12041305. 
    MWISP was sponsored by the National Key R\&D Program of China with grant 2017YFA0402701 and the CAS Key Research Program of Frontier Sciences with grant QYZDJ-SSW-SLH047. 
\end{acknowledgments}

\software{Astropy \citep{astropy2013, astropy2018}, Scipy \citep{scipy2020}, Matplotlib \citep{Hunter2007}}
    
\clearpage
\appendix
\renewcommand{\thefigure}{\Alph{section}\arabic{figure}}
\renewcommand{\theHfigure}{\Alph{section}\arabic{figure}}
\setcounter{figure}{0}

\section{The parameters of the DBSCAN algorithm}
The DBSCAN algorithm identifies a set of consecutive voxels (points) in the PPV cube as a molecular cloud.
The extracted voxels need to have $^{12}$CO line intensities above a certain threshold and 
connect with each other. There are three input parameters: cutoff, $\epsilon$, and MinPts.
The parameter of `cutoff' determines the line intensity threshold.
The $\epsilon$ and MinPts are for confining the connectivity of extracted structures. 
A core point is a point within the extracted contiguous structure, 
which have the adjacent points exceeding a number threshold in a certain radius. 
The `MinPts' determines the number threshold of adjacent points and the $\epsilon$ 
determines the radius of the adjacence.  
The border points of extracted structures are inside the $\epsilon$-neighborhood of core points, 
but do not include the `MinPts' neighbors in its $\epsilon$-neighborhood \citep{ester1996}. 
We adopt the cutoff = 2$\sigma$ ($\sigma$ is the rms noise, 
whose value is $\sim$ 0.5 K for $^{12}$CO line emission), 
MinPts=4, and $\epsilon$=1 in the DBSCAN algorithm for the identification of $^{12}$CO clouds, 
as suggested in \cite{Yan2020}. 
In addition, the post-selection criteria are also utilized to avoid noise contamination. 
That includes:(1) the total number of voxels in each extracted structure is larger than 16; 
(2) the peak intensity of extracted voxels is higher than the `cutoff' adding 3$\sigma$; 
(3) the angular area of the extracted structure is larger than one beam size (2$\times$2 pixels $\sim$ 1 arcmin$^{2}$); 
(4) the number of velocity channels needs to be larger than 3. 
The performance of different DBSCAN parameters on the extracted structures is presented in \cite{Yan2020, Yan2022}. 
The observational effects, including the finite angular resolution and sensitivity of the observed spectral lines data, 
on the extracted $^{12}$CO clouds also have been systematically investigated in \cite{Yan2022}.
In addition, the MC samples extracted by the DBSCAN algorithm also have been compared with 
those identified by other clustering algorithms, e.g., HDBSCAN and SCIMES, in \cite{Yan2022}.

The DBSACN parameters used for the extraction of $^{13}$CO structures 
are identical to the above parameters for $^{12}$CO clouds, 
except for the post-selection criteria of the peak intensities higher than the `cutoff' adding 2$\sigma$. 
In addition, the $\sigma$ is $\sim$ 0.25 K for $^{13}$CO line emission. 
We also compare the performance of three methods, including clipping, moment mask, and DBSCAN, on the 
extraction of $^{13}$CO structures in paper II. 

\section{The centroid velocities of $^{13}$CO line emission as systematic velocities of MCs}
\begin{figure*}[ht]
    \plotone{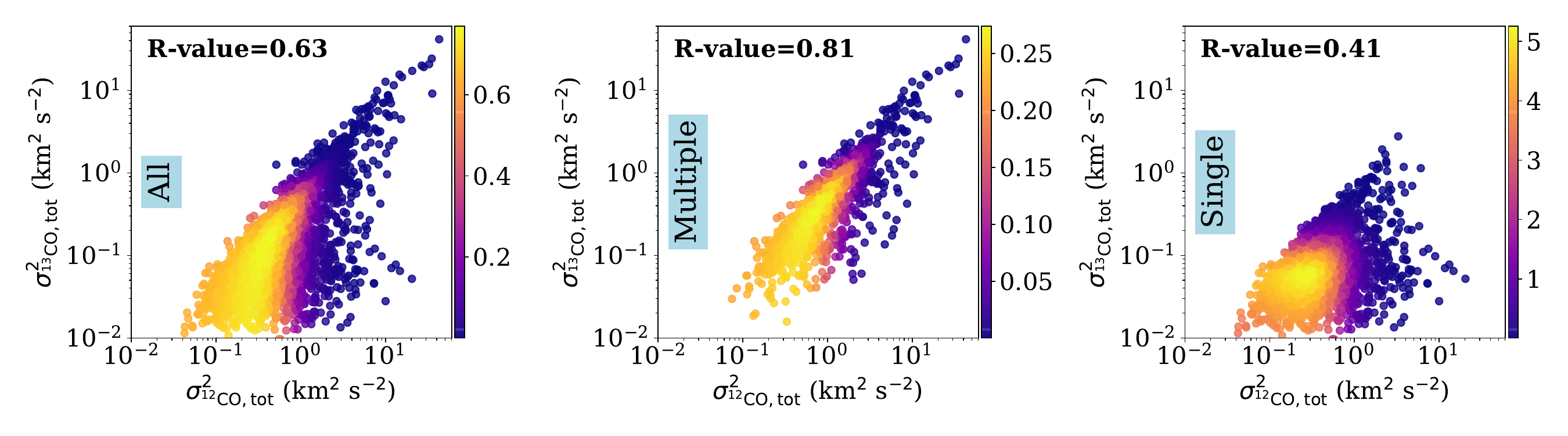} 
    \caption{The correlations between the $\sigma^{2}_{\rm ^{13}CO,tot}$ with the $\sigma^{2}_{\rm ^{12}CO,tot}$ 
    for the MC samples in the `All', `Multiple', and `Single' regimes, respectively. 
    The V$_{\rm sys}$ used in Eq.\ref{e:decom} is the V$_{\rm cen, ^{13}CO}$ of each MC. 
    Each dot in the panels represents a $^{12}$CO MC. 
    The colors on these dots represent the distribution of the probability density function(2D-PDF) of 
    their $^{12}$CO MCs. 
    The corresponding Spearman's rank correlation coefficient (R-value) is noted in each panel. \label{fig:f_corr2_13cen}}
\end{figure*}

\begin{figure*}[ht]
    \plotone{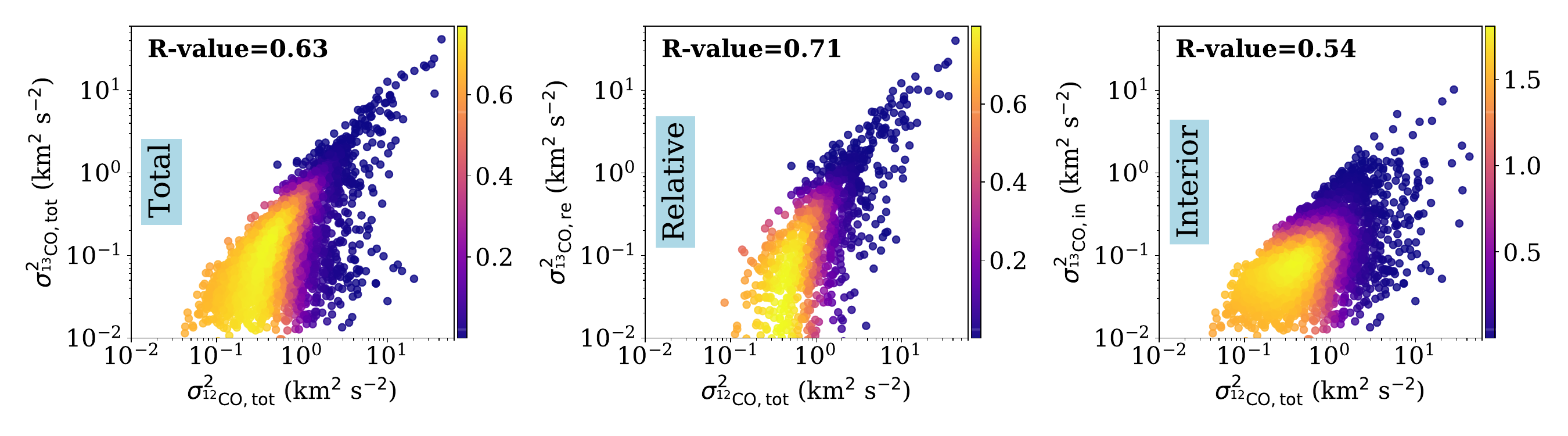} 
    \caption{The correlations between the $\sigma^{2}_{\rm ^{13}CO,tot}$, $\sigma^{2}_{\rm ^{13}CO,re}$, 
    and $\sigma^{2}_{\rm ^{13}CO,in}$ with $\sigma^{2}_{\rm ^{12}CO,tot}$, respectively, for all the samples. 
    The V$_{\rm sys}$ used in Eq.\ref{e:decom} is the V$_{\rm cen, ^{13}CO}$ of each MC. 
    Each dot in the panels represents a $^{12}$CO MC. 
    The colors on these dots represent the distribution of the probability density function(2D-PDF) of 
    their $^{12}$CO MCs.
    The corresponding Spearman's rank correlation coefficient (R-value) is noted in each panel. \label{fig:f_corr1_13cen}}
\end{figure*}

\begin{figure*}[ht]
    \plotone{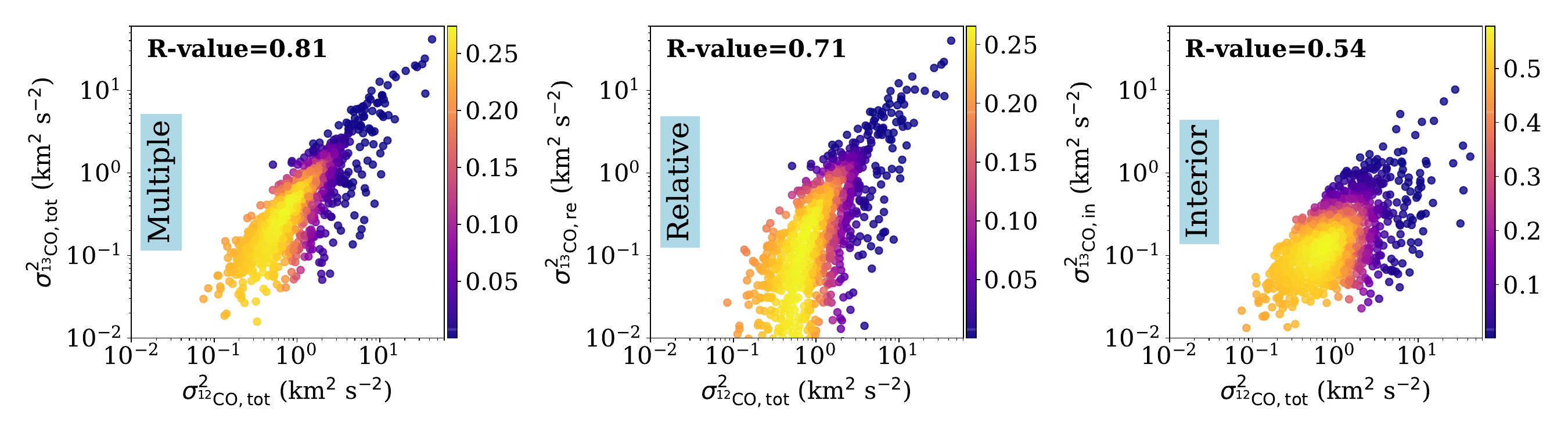} 
    \caption{The correlations between the $\sigma^{2}_{\rm ^{13}CO,tot}$, $\sigma^{2}_{\rm ^{13}CO,re}$, 
    and $\sigma^{2}_{\rm ^{13}CO,in}$ with $\sigma^{2}_{\rm ^{12}CO,tot}$, respectively, for the samples in the `multiple' regime. 
    The V$_{\rm sys}$ used in Eq.\ref{e:decom} is the V$_{\rm cen, ^{13}CO}$ of each MC. 
    Each dot in the panels represents a $^{12}$CO MC. 
    The colors on these dots represent the distribution of the probability density function(2D-PDF) of 
    their $^{12}$CO MCs. 
    The corresponding Spearman's rank correlation coefficients (R-value) is noted in each panel. \label{fig:f_corr3_13cen}}
\end{figure*}

The centroid velocities of $^{13}$CO line emission (V$_{\rm cen, ^{13}CO}$) are defined as the systematic velocities 
of MCs, which are used in Eq.\ref{e:decom} to calculate the velocity dispersions. 
We present the relations between the $\sigma^{2}_{\rm ^{13}CO,tot}$, $\sigma^{2}_{\rm ^{13}CO,re}$, 
and $\sigma^{2}_{\rm ^{13}CO,in}$ with the $\sigma^{2}_{\rm ^{12}CO,tot}$, for the MC samples. 
The MCs in the `All' regime can be divided into MCs in the `single' and `multiple' regimes. 
Figure \ref{fig:f_corr2_13cen} shows that relations between $\sigma^{2}_{\rm ^{13}CO,tot}$ and $\sigma^{2}_{\rm ^{12}CO,tot}$ 
for the MC samples in the `All', `Multiple', and `Single' regimes, respectively. 
Meanwhile, the $\sigma^{2}_{\rm ^{13}CO,tot}$ is decomposed into $\sigma^{2}_{\rm ^{13}CO,re}$ and $\sigma^{2}_{\rm ^{13}CO,in}$. 
Figure \ref{fig:f_corr1_13cen} presents the relations between the $\sigma^{2}_{\rm ^{13}CO,tot}$, $\sigma^{2}_{\rm ^{13}CO,re}$, $\sigma^{2}_{\rm ^{13}CO,in}$ 
with the $\sigma^{2}_{\rm ^{12}CO,tot}$ for all the MC samples, respectively. 
Also, Figure \ref{fig:f_corr3_13cen} demonstrates the same relations, but for the MCs in the multiples.


\section{Effects of optical depths on the relations}
\begin{figure}[ht]
    \plotone{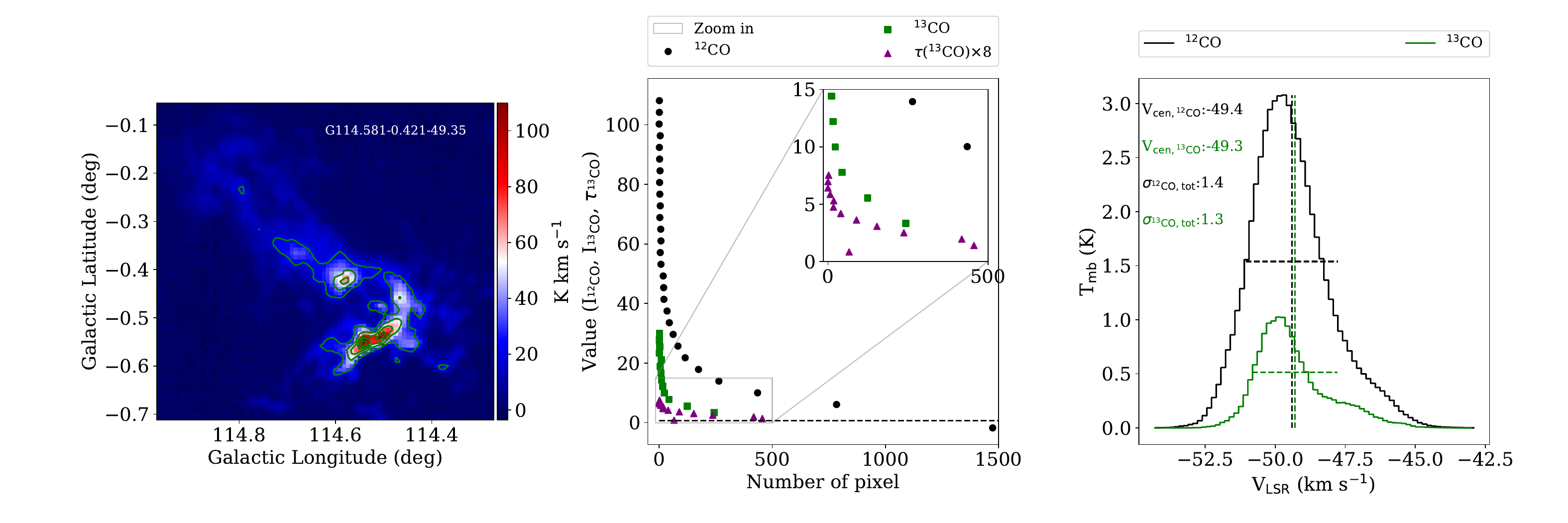} 
    \caption{\textbf{Left panel}: Distributions of velocity-integrated intensities of $^{12}$CO (I$_{\rm ^{12}CO}$, colormap) and 
    $^{13}$CO (I$_{\rm ^{13}CO}$, green contours) line emission for a MC named G114.581-0.421-49.35 in our catalog. 
    The green contours range from 10$\%$ to 90$\%$ stepped by 20$\%$ of the maximum value (31.1 K km s$^{-1}$). 
    \textbf{Middle panel}: Distributions of the pixel number in the intervals of values (I$_{\rm ^{12}CO}$, I$_{\rm ^{13}CO}$, $\tau_{\rm ^{13}CO}$). 
    \textbf{Right panel}: The averaged spectra for the extracted $^{12}$CO and $^{13}$CO lines emission within this MC. 
    The vertical black-dashed line delineates the centroid velocity of $^{12}$CO line emission (V$_{\rm cen,^{12}CO}$, km s$^{-1}$) 
    calculated as Eq. \ref{e:cen_12}. 
    The length of the horizontal black-dashed line shows the value of $\sqrt{8ln2}\times\sigma_{\rm ^{12}CO,tot}$, 
    which corresponds to the FWHM velocity-width of $^{12}$CO spectral line, and the $\sigma_{\rm ^{12}CO,tot}$ is calculated as Eq. \ref{e:decom}.
    The vertical green-dashed line delineates the centroid velocity of $^{13}$CO line emission (V$_{\rm cen,^{13}CO}$, km s$^{-1}$) 
    calculated as Eq. \ref{e:cen_13}. 
    The length of the horizontal green-dashed line shows the value of $\sqrt{8ln2}\times\sigma_{\rm ^{13}CO,tot}$, 
    which corresponds to the FWHM velocity-width of $^{13}$CO spectral line, and the $\sigma_{\rm ^{13}CO,tot}$ is calculated as Eq. \ref{e:decom}.\label{fig:ftau_trend1}}
\end{figure}

\begin{figure}[ht]
    \plotone{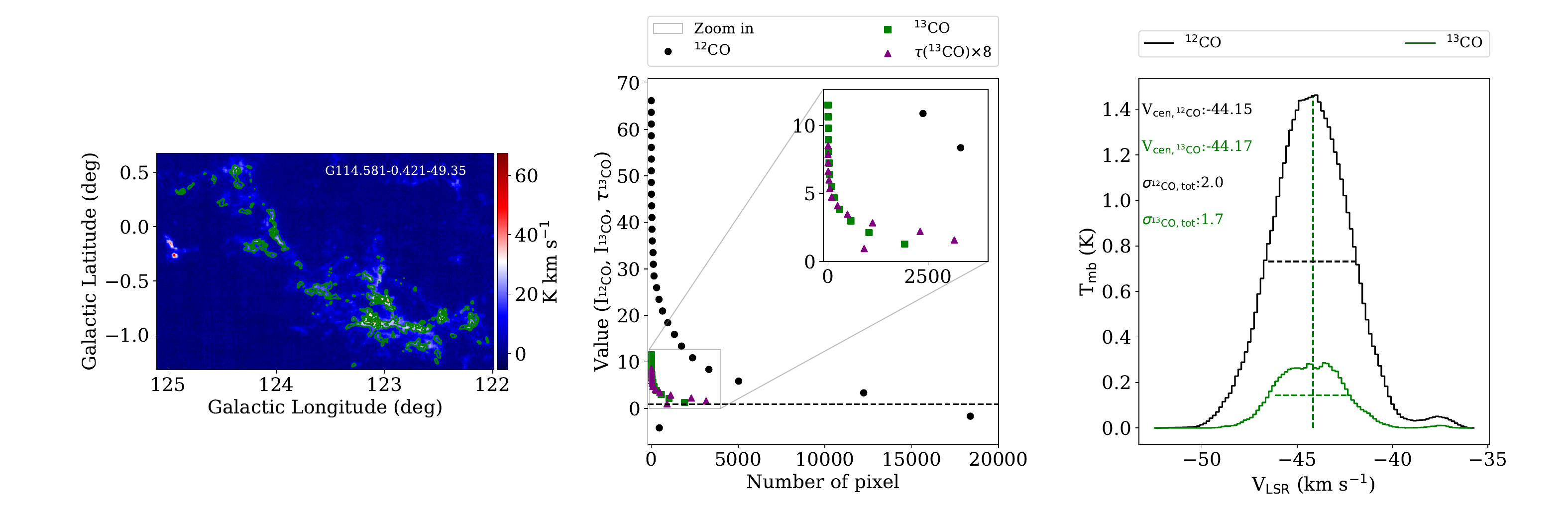}
    \caption{Same as Figure \ref{fig:ftau_trend1}, but for another MC named G123.291-0.539-44.15 in our catalog. 
    The green contours in the left panel range from 10$\%$ to 90$\%$ stepped by 20$\%$ of the maximum value (11.9 K km s$^{-1}$). \label{fig:ftau_trend2}} 

\end{figure}

In Figure \ref{fig:ftau_trend1} and \ref{fig:ftau_trend2}, 
we present the distributions of the velocity-integrated intensities of $^{12}$CO (I$_{\rm ^{12}CO}$) and $^{13}$CO (I$_{\rm ^{13}CO}$) emission for two MC samples 
in our catalog, and also the distributions of the pixel numbers in the intervals of the values (I$_{\rm ^{12}CO}$, I$_{\rm ^{13}CO}$, and $\tau_{\rm ^{13}CO}$). 
The values of $\tau_{\rm ^{13}CO}$ in the $^{13}$CO-bright regions of two MC samples are calculated as Eq. \ref{e:tau_13bri}. 

\begin{figure*}[ht]
    \plotone{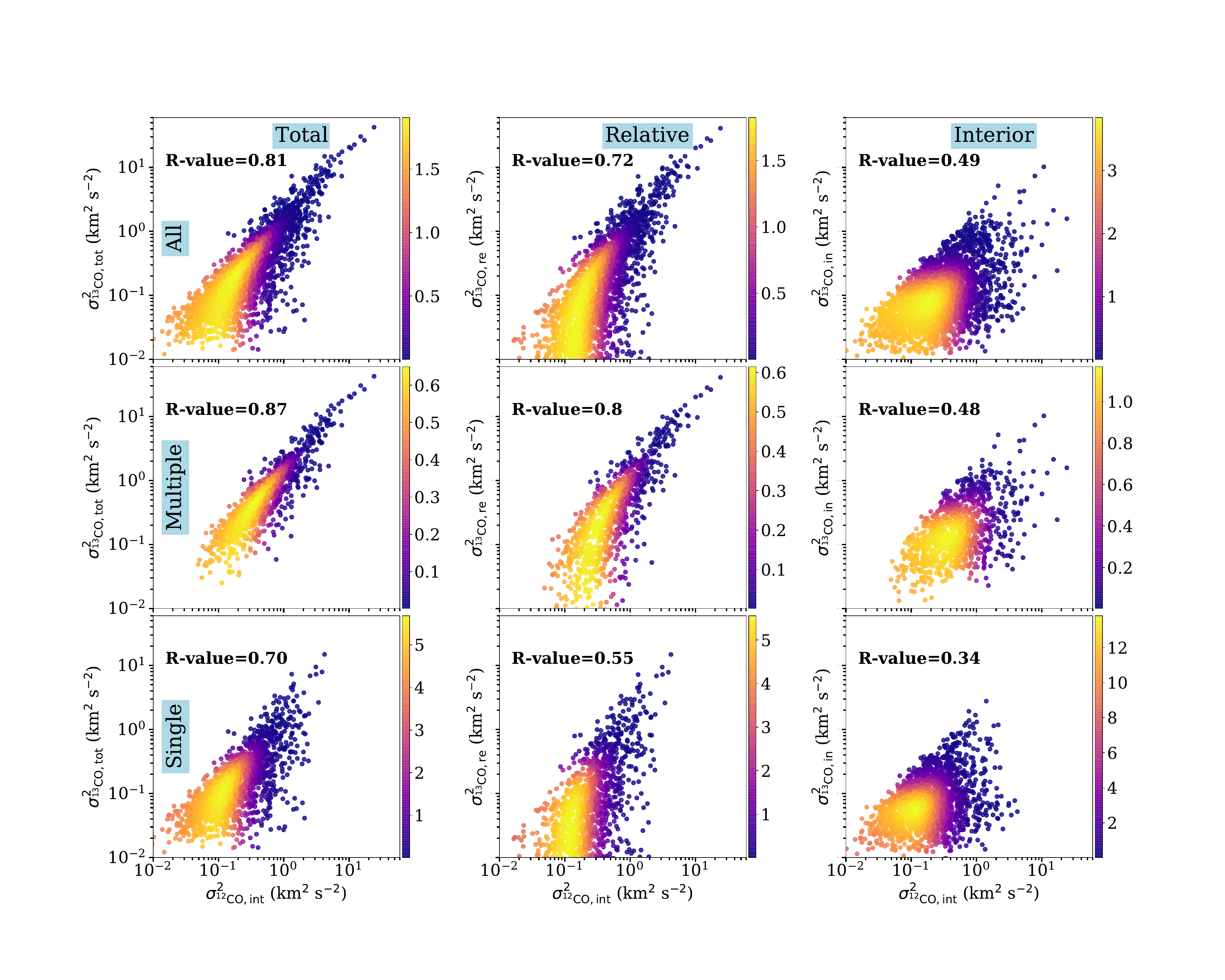}
    \caption{Relations between the velocity dispersion from different components($\sigma^{2}_{\rm ^{13}CO,tot}$, 
    $\sigma^{2}_{\rm ^{13}CO,re}$, and $\sigma^{2}_{\rm ^{13}CO,in}$) with the $\sigma^{2}_{\rm ^{12}CO,int}$, 
    whose values are revised the effect of optical depths as $\sigma_{\rm ^{12}CO,int}=\sigma_{\rm ^{12}CO,tot}/\bar\beta_{\tau}$.
    Their Spearman's rank correlation coefficients (R-value) are noted in each panel. \label{fig:ftau_rev}}
\end{figure*}

Figure \ref{fig:ftau_rev} shows the relations between the $\sigma^{2}_{\rm ^{13}CO,tot}$, 
$\sigma^{2}_{\rm ^{13}CO,re}$, and $\sigma^{2}_{\rm ^{13}CO,in}$ with $\sigma^{2}_{\rm ^{12}CO,int}$, 
where $\sigma_{\rm ^{12}CO,int}=\sigma_{\rm^{12}CO,tot}/\bar\beta_{\tau}$ to mitigate the effect of $^{12}$CO line opacity 
on the velocity dispersions of $^{12}$CO line emission.
    
\clearpage
\bibliography{vel_disper_13co.bib}{}
\bibliographystyle{aasjournal}


\end{document}